\documentclass[11pt]{article}
\usepackage{jheppub}
\usepackage{amsmath,amssymb,amsthm,mathtools,bm,mathrsfs,bbm}
\usepackage[dvipsnames]{xcolor}
\usepackage{comment}
\usepackage{hyperref}
\usepackage[utf8]{inputenc}
\usepackage[titletoc]{appendix}
\usepackage{tikz}
\usetikzlibrary{shapes,arrows,shadows}
\usepackage[colorinlistoftodos]{todonotes}
\usepackage{xspace}
\usepackage{etoolbox}
\usepackage{listofitems}
\usepackage{tikz-cd}
\usepackage{physics}
\usepackage[font={small,sf}]{caption, subcaption}
\usepackage[bottom]{footmisc}

\makeatletter
\def\@fpheader{\relax}
\makeatother

\theoremstyle{definition}
\newtheorem{definition}{Definition}[section]

\newcommand\be{\begin{equation}}
\newcommand\ee{\end{equation}}
\newcommand\beq{\begin{equation}}
\newcommand\eeq{\end{equation}}
\newcommand\bea{\begin{eqnarray}}
\newcommand\eea{\end{eqnarray}}
\newcommand\ba{\begin{array}}
\newcommand\ea{\end{array}}

\newcommand\eref[1]{(\ref{#1})}
\newcommand\bc{\begin{center}}
\newcommand\ec{\end{center}}

\newcommand{\Hkin}{\mathcal{H}^{\text{kin}}}
\newcommand{\Hgh}{\mathcal{H}^{\text{ghost}}}
\newcommand{\Hph}{\mathcal{H}^{\text{phys}}}
\newcommand{\bV}{\overline{V}}
\newcommand{\Vg}{V^\text{ghost}}
\newcommand{\gh}{\text{ghost}}
\newcommand{\omc}[1]{\overline{\mathcal{#1}}}
\newcommand{\tmc}[1]{\tilde{\mathcal{#1}}}
\newcommand{\bs}[1]{\boldsymbol{#1}}
\newcommand{\tbs}[1]{\tilde{\boldsymbol{#1}}}
\newcommand{\obs}[1]{\overline{\boldsymbol{#1}}}
\newcommand{\msn}{\mathsf{n}}
\newcommand{\bmsn}{\overline{\mathsf{n}}}
\newcommand{\msd}{\mathsf{d}}
\newcommand{\msC}{\mathsf{C}}

\newcommand{\brst}{\text{BRST}}
\renewcommand\comment[1]{}
\usepackage{blindtext}
\usepackage{enumitem}

\renewcommand\tilde{\widetilde}

\newcommand{\mc}[1]{\mathcal{#1}}
\newcommand{\nord}[1]{:\mathrel{#1}:}

\newcounter{descriptcount}
\newlist{enumdescript}{description}{1}
\setlist[enumdescript,1]{%
  before={\setcounter{descriptcount}{0}
          \renewcommand*\thedescriptcount{\arabic{descriptcount}}},
        font={\bfseries\stepcounter{descriptcount} \thedescriptcount.~}
}

\newtheoremstyle{indented}{3pt}{3pt}{\addtolength{\leftskip}{2.5em}}{}{\bfseries}{.}{.5em}{}

\theoremstyle{indented}

 \allowdisplaybreaks
 \numberwithin{equation}{section}

\title{\begin{center} Fortuity and Complexity in a Simple Quark Model \end{center}}

\author[*]{Jackson R.\ Fliss}
\author[\dagger,\ddagger]{\!\!, Vishnu Jejjala}
\author[\mathsection]{\!\!, Onkar Parrikar}

\affiliation[\,*]{Physique Th\'eoretique et Math\'ematique, Universit\'e Libre de Bruxelles \& International Solvay Institutes, CP 231, 1050 Bruxelles, Belgium}
\affiliation[\,\dagger]{Mandelstam Institute for Theoretical Physics, School of Physics, and NITheCS,\\ University of the Witwatersrand, Johannesburg, WITS 2050, South Africa}
\affiliation[\,\ddagger]{NSF AI Institute for Artificial Intelligence and Fundamental Interactions (IAIFI) and\\ Department of Physics, Northeastern University, Boston, MA 02115, USA}
\affiliation[\,\mathsection]{Department of Theoretical Physics, Tata Institute of Fundamental Research, Mumbai 400005, India}

\emailAdd{jackson.fliss@ulb.be}
\emailAdd{v.jejjala@wits.ac.za}
\emailAdd{parrikar@theory.tifr.res.in}

\abstract{
We observe and elaborate on a structural similarity between the categorization of monotone and fortuitous BPS operators in supersymmetric theories and gauge invariant quark operators in $SU(N_c)$ QCD.
Our designation of fortuity does not rely on supersymmetry and instead uses the BRST cohomology.
We argue that within this designation, baryon states are fortuitous while meson states are monotone.
We illustrate that in the Veneziano limit of large number of flavors and colors, this designation displays features resembling the fortuitous vs.\ monotone categorization of BPS operators, \textit{e.g.}, an exponential vs.\ polynomial dichotomy in the counting of operators.
We explore these ideas explicitly in a toy qubit model of quarks.
We further investigate the stabilizer R\'enyi entropy of meson and baryon states as a proxy for the complexity of classical simulation for these states. We show that all mesons display power law complexity and present evidence that typical baryons display super-exponential complexity in the Veneziano limit.

\vspace*\fill
\hspace*\fill{\textit{In Memoriam}}\\
\hspace*\fill{Robert G.\ Leigh}\\
\hspace*\fill{(1964 --- 2026)}
\vspace*\fill
}

\begin{document}

\maketitle
\pagebreak

\subsection*{Foreword}
{\it This paper is dedicated to Robert G.\ Leigh, who was the doctoral advisor, a collaborator, and a friend to each of the authors. To say that Rob made a lasting impact on each of our lives, both academically and personally, would be an understatement.
Through his hallmark inquisitiveness, through his careful and thoughtful explanations, through his high standards of rigor, and through his work ethic, Rob set an example of what it meant to be a theoretical physicist of the highest caliber.
Rob was effortlessly confident in expressing his ideas --- a confidence afforded by his natural insightfulness and his commitment to getting his hands dirty with computations --- and was unfazed by difficult problems and big questions.
He was a gifted teacher and mentor with a conviction that clarity of exposition was an intellectual obligation that we must fulfill as scientists.
As an advisor, Rob gave the impression that there is no sharp gap between faculty, postdocs, and students, that it didn’t matter who you were or where you came from: we were all curious researchers pursuing the same ends and if you put in the effort and had the ability, you could do great science.
Collaborating with Rob was one of the great pleasures in each of our careers.
He will be deeply missed by us and by his friends and colleagues in the theoretical physics community.}

\section{Introduction}\label{sec:intro}
AdS/CFT is best understood in the large $N$, 't Hooft limit where the boundary CFT is organized in a $\frac{1}{N}$ expansion, and correspondingly quantum effects are suppressed in the bulk.
However, holographic duality is envisaged as an exact equivalence of Hilbert spaces, and the importance of \emph{finite} $N$ effects on both sides of the duality has recently been emphasized.
In the particular context of BPS states in super Yang--Mills, a natural distinction arises between two sectors at large but finite $N$; these sectors correspondingly come with very different bulk interpretations. 

A BPS sector is \emph{monotone} if its cohomology classes persist as one increases the rank along a holographic large $N$ family of CFTs: concretely, the BPS operators/states can be organized into infinite sequences labeled by $N$ that remain BPS (in an appropriate supercharge cohomology) for all sufficiently large $N$.
By contrast, a BPS sector is \emph{fortuitous} if it consists of cohomology classes that exist only for a finite window of consecutive ranks and then disappear at larger $N$; this typically occurs because of finite $N$ relations (\textit{e.g.}, trace, or Cayley--Hamilton, relations in matrix algebras, or other rank-dependent constraints) which obstruct extending them.
The cohomological framework behind this distinction has an important prehistory in the $\frac1{16}$-BPS sector of $\mathcal N=4$ super-Yang--Mills: the finite $N$ counting problem was formulated in terms of local supercharge cohomology and compared with giant graviton and multigraviton states in~\cite{Grant:2008sk}, while the infinite $N$ multigraviton count and the associated puzzle for AdS$_5$ black hole entropy were sharpened in~\cite{Chang:2013fba}.
Explicit finite $N$ non-graviton cohomologies, together with their interpretation as black hole type states, were then found and developed in the $SU(2)$ theory in~\cite{Chang:2022mjp,Choi:2022caq}.
The monotone/fortuitous terminology and the associated ``holographic covering'' map were sharply formulated by Chang and Lin in~\cite{Chang:2024zqi}: the monotone sector behaves like a stable, perturbative (supergravity-like) subsector that can be consistently covered across $N$, whereas the fortuitous sector is intrinsically finite $N$ and therefore invisible in the large $N$ semiclassical bulk Hilbert space.
Instead they appear as non-perturbative objects in the bulk.

The proposed relevance to black holes is that monotone BPS states should be dual to smooth, horizonless supersymmetric geometries whose moduli spaces, once quantized, give the expected protected Hilbert spaces (with canonical examples including the $\frac12$-BPS geometries and their quantization on the AdS$_5$/CFT$_4$ side~\cite{Lin:2004nb}, and Lunin--Mathur microstate geometries on the AdS$_3$/CFT$_2$ side~\cite{Lunin:2001jy,Lunin:2002qf}).
Fortuitous states, on the other hand, are natural candidates for \emph{typical} black hole microstates which cannot be obtained from quantization of the classical gravitational phase space, but nevertheless contribute to the protected degeneracies responsible for BPS black hole entropy, in line with the modern superconformal index program for AdS black holes~\cite{Kinney:2005ej,Cabo-Bizet:2018ehj,Choi:2018hmj,Benini:2018ywd} and with the subsequent finite $N$ $\mathcal N=4$ SYM/non-graviton cohomology constructions~\cite{Chang:2022mjp,Choi:2022caq,Choi:2023znd,Choi:2023vdm,deMelloKoch:2024pcs,Gaikwad:2025ugk}.
Recent followups push on both sides of this picture: \textit{e.g.}, extensions/tests of fortuity in other models, including the SYK model~\cite{Chang:2024lxt}, the D1/D5 system~\cite{Chen:2024oqv,Chang:2025rqy}, supergravity~\cite{Hughes:2025car}, the ABJ vector model/higher-spin holography~\cite{Kim:2025vup}, ABJM theory~\cite{Belin:2025hsg}, and connections to chaotic dynamics and random matrices~\cite{Johnson:2026plw}.
Yet, there is still much to be understood about fortuity and black hole physics.
For instance, so far absent from the above discussions is the relation between fortuity and the exponentially large complexity expected of typical black microstates~\cite{Brown:2015lvg,Brown:2017jil,Brown:2019rox,Brandao:2019sgy}.
Additionally, given that key features of black hole microstates (such as chaotic dynamics and exponential degeneracy) are universal, it would be desirable to develop a framework for discussing finite $N$ effects outside the strict setting of BPS states and supersymmetric black holes.
Some progress was made in this direction in the context of matrix models in~\cite{deMelloKoch:2025ngs}.

In this paper, we observe the structural similarity between the cohomology of BPS states and the BRST cohomology of gauge invariant operators in gauge theories.
The core observation is simple.
Let us consider QCD with color gauge group $SU(N_c)$ with $N_f$ flavors of quarks and anti-quarks in the fundamental and antifundamental representations and focus on gauge invariant operators that are bound states of the fermions.
The simplest gauge invariant operators are the mesons, which pair a quark with an anti-quark and contract color indices:\footnote{Unless specifically noted, we will use Einstein summation notation on color indices, where repeated upper/lower index pairs are summed over. This will also be true of Lie algebra indices. Because we will consider flavor symmetry as a global symmetry, physical quantities can possess free, and possibly repeated, flavor indices. As such we will explicitly denote flavor index summations.}
\be
M_{ij} = q^{a}_i \overline{q}_{j,a} ~, \qquad a=1,\ldots,N_c ~. \label{eq:meson}
\ee
Since the flavor indices run from $i,j=1,\ldots,N_f$, there are $N_f^2$ distinct mesons.
If we increment the number of colors, the number of mesons does not change.
Mesons are the analogues of monotone operators.
We also have baryon operators constructed from taking the combinations
\be
B_{i_1,\ldots,i_{N_c}} = \epsilon_{a_1,\ldots,a_{N_c}} q^{a_1}_{i_1} \ldots q^{a_{N_c}}_{i_{N_c}} ~. \label{eq:baryon}
\ee
The anti-symmetrization of the color indices requires exactly $N_c$ quarks to construct a gauge invariant singlet of $SU(N_c)$.
The expression is symmetric in the flavor indices due to the Pauli statistics of the quarks. Similarly, we have anti-baryons
\be
\overline{B}_{j_1,\ldots,j_{N_c}} = \epsilon^{a_1,\ldots,a_{N_c}} \overline{q}_{j_1,a_1} \ldots \overline{q}_{j_{N_c},a_{N_c}} ~. \label{eq:anti-baryon}
\ee
If we send $N_c\to N_c+1$, the baryons and anti-baryons are structurally different in the new theory: we need to add an additional quark (or anti-quark) to construct a gauge invariant operator because the $\epsilon$-symbol for $SU(N_c+1)$ has an extra color index to contract.
The baryons from the $SU(N_c)$ theory are no longer gauge invariant in the $SU(N_c+1)$ theory.
Baryons are the analogues of fortuitous operators. In this paper, we will formalize this argument through an appropriate covering map \emph{\`a la}~\cite{Chang:2024zqi,Chang:2024lxt} based on BRST cohomology. We pause to note that the Chang--Lin taxonomy~\cite{Chang:2024zqi} is formulated in the supercharge cohomology of protected BPS sectors, which is {\it dynamical} and protected; BRST gauge cohomology, on the other hand, is a {\it kinematical} implementation of gauge invariance.
Our construction is therefore best understood as a structural analogue of fortuity, not as evidence for the same physical mechanism that governs typical BPS black hole microstates.

The enumeration of the gauge invariant operators in QCD is likewise similar to the counting of monotone and fortuitous BPS states.
We have seen that the number of mesons grows quadratically in $N_f$ for any $N_c$.
What about the baryons?
For simplicity, let us ignore spin and orbital excitations and just count flavor multiplicities of the simplest color singlet operators as the extra quantum numbers do not affect the crucial point.
Quarks are fermions, but the color wavefunction is totally antisymmetric, and for the simplest $s$-wave/spin-symmetric ground states are totally symmetric in their flavor indices.
The number of totally symmetric tensors of rank $N_c$ on an $N_f$-dimensional space is the number of ways of choosing a multiset of $N_c$ flavors from $N_f$ possibilities.
Thus, we calculate
\be
\text{Number of baryons} = \dim \text{Sym}^{N_c} (\mathbb{C}^{N_f}) = {{N_f + N_c - 1}\choose N_c} ~. \label{eq:countingbaryons}
\ee

When $N_f = N_c = 3$, we can apply these formul\ae\ to the familiar hadrons~\cite{Gell-Mann:1961omu,Neeman:1961jhl}.
We have nine mesons, $\bm{3}\otimes \overline{\bm{3}} = \bm{8}\oplus \bm{1}$.
The flavor octet consists of $\pi^+, \pi^0, \pi^-, K^+, K^0, \overline{K}^0, K^-, \eta_8$.
The flavor singlet is $\eta_1$, and the physical $\eta$ and $\eta'$ mesons arise from the mixing between $\eta_1$ and $\eta_8$.
In the baryonic sector, we count
\be
uuu,\, ddd,\, sss,\, uud,\, uus,\, udd,\, uss,\, dds,\, dss,\, uds
\ee
as the ten monomials of degree three in three variables, which is indeed $5\choose 3$; these hadrons comprise the $J=\frac32$ decuplet of baryons.
This enumeration ignores the spin representation of the quarks.
There are additionally the $J=\frac12$ states, which include the proton and the neutron, which come in an octet ({\it i.e.}, the eightfold way); these states arise from a different Young diagram corresponding to a mixed symmetric representation of $SU(2)_\text{spin}\times SU(3)_f$.
Including these states sends $N_f\to 2N_f$ in~\eref{eq:countingbaryons}.\footnote{
This is strictly true in the simplest maximally symmetric spin-flavor sector.
For other fully local QCD operators with specified Lorentz and Dirac structure, additional Fierz, equation of motion, and integration by parts relations must be imposed.}
Thus in total, we have ${8\choose 3} = \mathbf{56} = (\mathbf{8}, \mathbf{2}) \oplus (\mathbf{10}, \mathbf{4})$, explaining the decomposition into the flavor octet $\times$ spin doublet and flavor decuplet $\times$ spin quartet.

We will be interested in the enumeration of these operators in the Veneziano double scaling limit: \textit{i.e.}, large $N_f$, large $N_c$, with $\xi := N_f/N_c$ fixed~\cite{Veneziano:1976wm}.
There are several reasons that make this limit the relevant ``large $N$'' limit for quark operators.
This limit retains the simplifying power of large $N_c$ methods while keeping the flavor sector dynamically important.
In the strict 't~Hooft limit with fixed $N_f$~\cite{tHooft:1973alw}, quark loops are parametrically suppressed relative to gluon contributions, so the theory is dominated by pure gauge dynamics and many effects associated with dynamical quarks are pushed to subleading order.
By contrast, in the Veneziano limit, quark loops survive at leading order, mesons and baryons remain numerous, and the counting of states reflects the competition between color and flavor in a way that is closer to real QCD.
For small $\xi$, the theory confines, whereas for larger $\xi$ within a conformal window, the theory can reach an infrared interacting (Banks--Zaks) fixed point where it becomes scale invariant~\cite{Banks:1981nn,Appelquist:1996dq}.
Working in the Veneziano limit in holographic QCD, the large number of flavors means the associated flavor branes have enough tension to deform the spacetime geometry.
This backreaction is necessary to holographically model phenomena like the running of the coupling constant in the presence of dynamical matter~\cite{Bigazzi:2005md,Jarvinen:2011qe}.

In the Veneziano limit, using Stirling's approximation on~\eref{eq:countingbaryons}, we see that the number of baryon operators grows exponentially with $N_c$~\cite{Veneziano:1976wm,Manohar:1998xv}:
\be
    \text{Number of baryons} \approx e^{\zeta N_c} ~, \qquad \zeta = (1+\xi) \log(1+\xi) - \xi \log\xi ~.
\ee
We contrast this with the number of mesons which only grows polynomially, $\xi^2 N_c^2$.
The behavior in this limit is parametrically the same even when we are more careful about treating spin.
The natural monotone sector is the sector generated without $\epsilon$-tensors.
The fortuitous sector is the sector that genuinely needs an $\epsilon$-tensor whose rank depends on $N_c$.
In analogy to BPS fortuitous states, the baryons dominate the spectrum of gauge invariant operators in this toy model.\footnote{
By ``toy model,'' we mean that we work in a simplified quark only subsector of the gauge theory.
The comparison is made at fixed minimal baryon number/constituent number.
Real QCD operator counting is considerably more involved: for instance, we have neglected the gluons entirely.
The additional complications are precisely why Hilbert series and operator basis methods are used in effective field theory approaches to the counting problem~\cite{Lehman:2015via,Henning:2015daa}.}

In the following sections, we elaborate on this similarity.
In Section~\ref{sec:qubit}, we construct a qubit model of quarks and formalize the definition of monotone and fortuitous within its BRST cohomology.
Our model is extremely simple --- it elides all of the important quantum field theoretic subtleties associated to real QCD, plus other features such as spin, gluon states, interactions, or even dynamics for that matter --- its primary role is to highlight the kinematic features of fortuity and monotony, \textit{i.e.}, those associated directly to $SU(N_c)$ representation theory.
Regardless of these simplifications, we expect the core lessons from this kinematic form of monotony and fortuity to extend to the fermionic gauge invariant operators of QCD naturally.

A second goal of this article is to explore the idea that monotonous states are \emph{low complexity}, \textit{i.e.}, they admit a simple, semiclassical description in the bulk, while fortuitous states are \emph{high complexity}, and do not admit a simple, semiclassical description. A primary benefit of our simplified quark model is that it will allow us to assign a concrete notion of state complexity to our monotone and fortuitous states.
The relevant notion of complexity we will use is that of \emph{stabilizer complexity}, or intuitively the complexity of classical simulation, and is rooted in the resource theory of {\it magic states} within the context of stabilizer based quantum computation, which we review in Section~\ref{sec:sre} (see~\cite{White:2020zoz, Cao:2023mzo, Cao:2024nrx, Basu:2024tgg, Basu:2025mmm, Basu:2025uxw, Malvimat:2026oqf, Bettaque:2026vpl} for some recent work on stabilizer complexity).\footnote{
There is also a related literature on magic in constrained Hilbert spaces.  In particular, non-stabilizerness has been studied in discrete lattice gauge theories~\cite{Esposito:2025lgtmagic}, in the $SU(2)$ gauge invariant intertwiner space of quantum tetrahedra~\cite{Cepollaro:2024tetrahedra}, and more generally for subspaces and their embeddings, including symmetry constrained examples~\cite{Cepollaro:2025subspaces}.}
In our qubit toy model, we study a natural measure of stabilizer complexity called the {\it stabilizer R\'enyi entropy}~\cite{Leone:2021rzd} with respect to the Fock basis, and show that from this point of view the complexity of mesons scales at most polynomially in $N_c$, while that of typical baryon states scales as $e^{N_c\log N_c}$ in the Veneziano limit.

Lastly, in Section~\ref{sec:disc}, we conclude with interesting takeaway messages, some subtleties arising from the simplicity of our model, and finally some speculative parallels between gauge invariant operators in QCD and black hole microstates.

\section{Fortuity in a qubit model of quarks}\label{sec:qubit}

Let us introduce our toy model of ``dynamics-less'' and ``spin-less'' quarks and anti-quarks.
We will treat quarks and anti-quarks as separate sets of complex fermions acting on a qubit Hilbert space.
We denote
\beq\label{eq:qbHS}
    V^a_{i}=\text{span}_{\mathbb C}\Big\{\ket{0}^a_{i},\ket{1}^a_{i}\Big\}~,\qquad \bV_{a,i}=\text{span}_{\mathbb C}\Big\{\ket{\overline 0}_{a,i},\ket{\overline 1}_{a,i}\Big\}~,
\eeq
as a single quark and single anti-quark Hilbert spaces, respectively.
We have labeled these spaces by color and flavor indices which run from $a=1,\ldots, N_c$ and $i=1,\ldots, N_f$, respectively.
Acting on these spaces are quark and anti-quark operators $\{q_{a,i}\}$ and $\{\overline q^a_i\}$ satisfying\footnote{
As an index rule we denote $(q^a_i)^\dagger\equiv q^\dagger_{a,i}$. 
The reason for this is that $q$ and $q^\dagger$ transform in conjugate representations as will become clear.}
\beq
    q^a_i\ket{0}^a_i=\ket{1}^a_i~,\qquad q^{\dagger}_{a,i}\ket{1}^a_i=\ket{0}^a_i~,\qquad q^a_i\ket{1}^a_i=q^{\dagger}_{a,i}\ket{0}^a_i=0~,
\eeq
and similarly for $\{\overline q_{a,i}\}$ on the $\bV_{a,i}$ spaces.
Note that $q^a_i$ is defined to only act on the $V^a_i$ with a fixed $(a,i)$.
That is, quarks of different color and flavor indices are decoupled and satisfy anti-commutators 
\beq
    \{q^a_i,q^{\dagger}_{b,j}\}=\delta^a_b\delta_{ij}~,\qquad \{\overline q_{a,i},\overline q^{\dagger b}_{j}\}=\delta^b_a\delta_{ij}~,
\eeq
with all others zero.

We will denote the ``kinematic'' Hilbert space at fixed $N_c$ and $N_f$ as
\beq
    \Hkin_{N_c,N_f}=\bigotimes_{a=1}^{N_c}\bigotimes_{i=1}^{N_f}\Big(V^a_i\otimes \bV_{a,i}\Big)~.
\eeq
It is useful to grade this Hilbert space into total quark and anti-quark number as
\beq
    \Hkin_{N_c,N_f}=\Big(\bigoplus_{n=0}^{N_cN_f}\mc H^{(n)}\Big)\otimes\Big(\bigoplus_{\bar n=0}^{N_cN_f}\omc H^{(\bar n)}\Big)~.
\eeq
$\mc H^{(n)}$ and $\omc H^{(\bar n)}$ are eigenspaces of the quark and anti-quark number operators
\beq
    \msn=\sum_{i}q^a_iq^{\dagger}_{a,i}~,\qquad \bmsn=\sum_{i}\overline q_{a,i}\overline q^{\dagger a}_{i}~,
\eeq
with eigenvalues $n$ and $\bar n$ respectively. Of particular note are the zero and one (anti-)quark sectors.
Within these sectors we will denote
\beq
    \ket{\bs 0}\equiv \bigotimes_{a,i}\ket{0}^a_{i}\in\mc H^{(0)}~,\qquad \ket{\bs 1^a_{i}}\equiv q^a_{i}\ket{\bs 0}=\ket{0}\otimes\ldots\ket{1}^a_{i}\otimes\ldots\ket{0}\in\mc H^{(1)}~,
\eeq
with similar definitions for $\ket{\obs{0}}$ and $\ket{\obs{1}_{a,i}}$.
We define an $SU(N_c)$ action on these two sectors with $\mc H^{(0)}$ furnishing the trivial representation and $\mc H^{(1)}$ furnishing the fundamental representation.
The spaces $\mc H^{(n>1)}$ then furnish tensor product representations of the fundamental.
To be specific, for $U\in SU(N_c)$, the action is
\beq\label{eq:SUgroupact1}
    U\ket{\bs 0}= \ket{\bs 0}~,\qquad U\ket{\bs 1^a_{i}}= {U^a}_b\,|\bs 1^b_{i}\rangle~.
\eeq
Similarly, the anti-quark spaces lie in tensor products of the anti-fundamental representation:
\beq\label{eq:SUgroupact2}
U\ket{\obs 0}= \ket{\obs 0}~,\qquad U\ket{\obs 1_{a,i}}= {U^{\dagger b}}_a\,\ket{\obs{1}_{b,i}}~.
\eeq
The corresponding group action on the quark operators is
\beq
    U:q^a_{i}\rightarrow {U^a}_bq^b_i~,\qquad U:\overline q_{a,i}\rightarrow \overline q_{b,i}{U^{\dagger b}}_a~.
\eeq
Note that this implies that the quark annihilation operator lies in the anti-fundamental:
\beq
    U:q^\dagger_{a,i}\rightarrow ({U^a}_b)^\ast (q^b_i)^\dagger = q^\dagger_{b,i}{U^{\dagger b}}_a~.
\eeq
The above action on quarks is generated by the operator: 
\beq\label{eq:Gexplicit}
    G_A=\sum_{i}\nord{q^\dagger_i (T_A)q_{i}}-\sum_{i}\nord{\overline q_i (T_A)\overline q_{i}^\dagger}~,
\eeq
which satisfies
\beq\label{eq:GAcomms}
    [\hat G_A,q^a_{i}]={(T_A)^a}_bq^b_i~,\qquad [\hat G_A,\overline q_{a,i}]=-\overline q_{b,i}{(T_A)^b}_a~,
\eeq
and where $T_A$ are Hermitian and traceless generators of $SU(N_c)$ with structure constants\footnote{
Our $SU(N_c)$ conventions can be found in Appendix~\ref{app:fabc}.}
${f^C}_{AB}$:
\beq\label{eq:SUNalg}
    [T_A,T_B]=i{f^C}_{AB}T_C~,\qquad \Tr_\text{fund}\left(T_AT_B\right)=\frac{1}{2}\delta_{AB}~.
\eeq
We additionally note that the generators~\eqref{eq:Gexplicit} commute with both quark and anti-quark number operators
\beq\label{eq:Gkillsn}
    [G_A,\msn]=[G_A,\bmsn]=0~.
\eeq
In~\eqref{eq:Gexplicit}, we have introduced a ``normal ordering'' that places all $q^\dagger$'s to the right and $q$'s to the left\footnote{
Explicitly,
\[
    G_A=\sum_{i}{(T_A)^a}_b q^b_{i}\,q^\dagger_{a,i}-\sum_i{(T_A)^a}_b\overline q_{a,i}\overline q^{\dagger b}_i~.
\]}
such that the zero particle sector is annihilated:
\beq\label{eq:Gkillsvac}
    \hat G_A\ket{\bs 0}=\hat G_A\ket{\obs 0}=0~.
\eeq
It then follows from~\eqref{eq:GAcomms} that
\beq
    \hat G_A\ket{\bs 1^a_{i}}={(T_A)^a}_b|\bs 1^b_i\rangle~,\qquad G_A\ket{\obs 1_{a,i}}=-{(T_A)^b}_a\ket{\obs 1_{b,i}}~,
\eeq
and the group actions~\eqref{eq:SUgroupact1} and~\eqref{eq:SUgroupact2} are obtained through exponentiation, \textit{e.g.}, $\exp (i\lambda^AT_A)$ with $\lambda^A\in\mathbb R$.

We wish to gauge this $SU(N_c)$ symmetry, and so we must identify the physical, gauge-invariant subspace of $\Hkin$.
To do this, we will construct a set of BRST charges by introducing a set of $N_c^2-1$ ghost fields $\{c^A\}$ (with their Hermitian conjugate $\{c_A^{\dagger}\}$).
In order to accommodate this ghost space, we will first extend $\Hkin_{N_c,N_f}$ to
\beq
    \tmc H_{N_c,N_f}\equiv \Hkin_{N_c,N_f}\otimes \Hgh_{N_c}~,\qquad \Hgh_{N_c}\equiv \bigotimes_{A=1}^{N_c^2-1}\Vg_A~,
\eeq
where $\Vg_A$ is the two-state fermion Hilbert space,
\beq
    \Vg_A=\text{span}_{\mathbb C}\Big\{\ket{0}_A^\gh,\ket{1}_A^\gh\Big\}~.
\eeq
The ghost Hilbert space can be graded under the eigenvalue of $\mathsf{n}^\text{ghost}=c^Ac_A^\dagger$, which we will call the ``ghost number'' and denote by $\mathsf{g}$:
\beq
    \Hgh_{N_c}=\bigoplus_{\mathsf{g}}\Hgh_{N_c,(\mathsf{g})}~.
\eeq
We will denote then tensor products of the quark, anti-quark, and ghost vacua as
\beq
    |\tbs{0}\rangle:=\ket{\bs{0}}\otimes\ket{\obs{0}}\otimes\ket{0}_\text{ghost}~.
\eeq
Introducing a BRST charge
\beq\label{eq:BRSTQ}
    Q_\brst\equiv c^A\hat G_A+\frac{i}{2}{f^A}_{BC}\,c^Bc^Cc_A^{\dagger}~,
\eeq 
then defines a BRST differential, $\bs{s}$, as
\beq
    \bs{s}(\cdot):=ic^A\delta_A(\cdot)\equiv i[Q_\brst,\cdot]~,
\eeq
for fermion-even operators and 
\beq
    \bs{s}(\cdot)\equiv i\{Q_\brst,\cdot\}~,
\eeq
for fermion-odd operators. $Q_\brst$ is fermionic and has ghost number 1.
We have normal ordered it so the extended vacuum is annihilated by both $Q_\brst$ and $Q^\dagger_\brst$:
\beq
    Q_\brst|\tbs{0}\rangle=Q^\dagger_\brst|\tbs{0}\rangle=0~.
\eeq
Additionally, due to the ghost variation
\beq
    \{Q_\brst,c^A\}=\frac{i}{2}{f^A}_{BC}c^Bc^C~,
\eeq
the BRST differential is nilpotent (see Appendix~\ref{app:BRST} for details)
\beq
    \bs{s}^2(\cdot)=0~.
\eeq
This allows us to define a cohomology at ghost number $\mathsf{g}$ as 
\beq
    \mathsf{H}^{(\mathsf{g})}:=\Big\{\ket{\psi}\in\Hkin_{N_c,N_f}\otimes \Hgh_{N_c,(\mathsf{g})}\Big|Q_\brst\ket{\psi}=0\Big\}\Big/\Big\{Q_\brst\ket{\phi}~,~\ket{\phi}\in \Hkin_{N_c,N_f}\otimes \Hgh_{N_c,(\mathsf{g-1})}\Big\}~.
\eeq

\subsection{Gauge invariant states}
We identify physical, gauge-invariant states as the ghost number-$0$ BRST cohomology.
Ghost number-$0$ states cannot be BRST exact; it is both sufficient and necessary that such states are annihilated by $Q_\brst$:
\beq
    \Hph_{N_c,N_f}:=\mathsf{H^{(0)}}=\Big\{\ket{\psi}\in\Hkin_{N_c,N_f}\otimes \Hgh_{N_c,(\mathsf{0})}\Big|Q_\brst\ket{\psi}=0\Big\}~.
\eeq
Explicitly, on ghost vacuum states,
\beq
Q_\brst \ket{\psi} \otimes \ket{0}_\gh =0 \quad \Longleftrightarrow \quad G_A = 0\ \ \forall A~.
\eeq
There are two important classes of gauge invariant states that will be the focus of this paper.
The first are \emph{meson states} formed out of quark--anti-quark bilinears:
\beq\label{eq:qbmesstate}
    \ket{M_{ij}}:=\frac{1}{\sqrt{N_c}}\overline q_{a,i}\,q^a_j\ket{\tbs{0}}~.
\eeq
The second are \emph{baryon} and \emph{anti-baryon states} formed out of contracting quarks and anti-quarks with epsilon tensors (respectively).
As mentioned in the Section~\ref{sec:intro}, baryons are completely symmetric in their flavor indices.
Thus they are uniquely labeled by a partition of $N_f$ into $N_c$ parts,
\beq
    \{m_i\}_{i=1,\ldots, N_f}~,\qquad \sum_{i=1}^{N_f}m_i=N_c~;
\eeq
that is, $m_i$ counts how many times the particular flavor index $i$ appears in $B_{i_1\ldots i_{N_c}}$.
The normalized baryon state is given by
\beq\label{eq:qbbarstate}
    \ket{B_{i_1i_2\ldots i_{N_c}}}:=\sqrt{\frac{1}{N_c!\prod_{i=1}^{N_f}(m_i!)}}\epsilon_{a_1a_2\ldots a_{N_c}}q^{a_1}_{i_1}q^{a_2}_{i_2}\ldots q^{a_{N_c}}_{i_{N_c}}|\tbs{0}\rangle~,
\eeq
and analogously for $\ket{\overline{B}_{i_1,i_2,\ldots,i_{N_c}}}$.

It is easy to verify that these states are annihilated by $Q_\brst$ as
\begin{align}
    Q_\brst\ket{M_{ij}}&=\frac{i}{\sqrt{N_c}}c^A\delta_{A}(\overline q_{a,i}\,q^a_j)|\tbs{0}\rangle=0~,\nonumber\\
    Q_\brst\ket{B_{i_1i_2\ldots i_{N_c}}}&=i\sqrt{\frac{1}{N_c!\prod_{i}(m_i!)}}c^A\delta_{A}\left(\epsilon_{a_1a_2\ldots a_{N_c}}q^{a_1}_{i_1}q^{a_2}_{i_2}\ldots q^{a_{N_c}}_{i_{N_c}}\right)|\tbs{0}\rangle=0~,\nonumber\\
    Q_\brst\ket{\overline B_{i_1i_2\ldots i_{N_c}}}&=i\sqrt{\frac{1}{N_c!\prod_{i}(m_i!)}}c^A\delta_{A}\left(\epsilon_{a_1a_2\ldots a_{N_c}}\overline q_{a_1,i_1}\overline q_{a_2,i_2}\ldots \overline q_{a_{N_c},i_{N_c}}\right)|\tbs{0}\rangle=0~,
\end{align}
the latter two because of tracelessness of $T_A$.

\subsection{The covering space}
We have a construction of the kinematic Hilbert space for each $N_c$ and have identified a Hilbert space of physical states as the zero ghost cohomology class of $Q_\brst$.
We can now ask what becomes of states as we incrementally increase $N_c$. To do so we pass to the covering space, \textit{i.e.}, the collection of BRST cohomology classes over all $N_c$ and compatible embeddings which we will construct sequentially.
For convenience, we will eliminate notating the explicit dependence of various quantities on flavor, although, as outlined in Section~\ref{sec:intro}, we will be interested in letting $N_f$ scale with $N_c$.
The kinematic Hilbert spaces admit an incremental embedding scheme
\beq\label{eq:Hkinembed}
    \Hkin_{N_c}\hookrightarrow\Hkin_{N_c+1}~,
\eeq
where we can identify $\Hkin_{N_c}$ as the image of the projector $\pi_{N_c+1}$ as
\beq
    \pi_{N_c+1}\Hkin_{N_c+1}=\Hkin_{N_c}\otimes\ket{\omega_{N_c+1}}~,\qquad \ket{\omega_{N_c+1}}=\bigotimes_{i=1}^{N_f}\ket{0}_{N_c+1,i}\otimes\bigotimes_{j=1}^{N_f}\ket{\overline 0}_{N_c+1,j}~.
\eeq
That is we just put ${(N_c+1)}^\text{th}$ (anti-)quark in its unoccupied state. 

We also need to state what to do with the ghost sector.
Let us order the generators of $SU(N_c+1)$ according to the sequential canonical embeddings
\beq\label{eq:suNembed1}
    1\hookrightarrow SU(2)\hookrightarrow SU(3)\hookrightarrow\ldots\hookrightarrow SU(N_c)\hookrightarrow SU(N_c+1)
\eeq
that sends each $U\in SU(N_c)$ to 
\beq\label{eq:suNembed2}
    U\rightarrow \left(\begin{array}{cc}U & 0\\ 0 &1\end{array}\right)\in SU(N_c+1)~.
\eeq
Under this ordering, the generators $\{T_{A}\}_{A<N_c^2}$ of $SU(N_c+1)$ all have zeros in their $(N_c+1)^\text{th}$ row and column and generate an $\mathfrak{su}(N_c)$ subalgebra of $\mathfrak{su}(N_c+1)$.

We then define the embedding and corresponding projector\footnote{
This is really the embedding $\widetilde{\mathcal H}_{N_c,N_f} \hookrightarrow \widetilde{\mathcal H}_{N_c+1,N_f}$.
That is to say, for the purposes of the cohomological covering map, $N_f$ is held fixed.
The Veneziano limit we use later should be understood as a choice of a sequence of theories and ensembles of baryon states; it is not part of the definition of the rank covering map.
However, the result is robust.
If we define a flavor tower as well as a color tower, for example, by choosing an infinite flavor space and setting $N_f(N_c)=\lfloor \xi N_c\rfloor$, then we can define
$\widetilde{\mathcal H}_{N_c,N_f(N_c)} \hookrightarrow \widetilde{\mathcal H}_{N_c+1,N_f(N_c+1)}$
by adding both the new color qubits and any newly activated flavor qubits in their vacua. Existing mesons built from old flavors remain monotone.
Existing baryons still fail to embed as baryons of the larger color group because the obstruction is a color-rank obstruction, not a lack of flavors.}
\beq
    \tmc H_{N_c}\hookrightarrow\tmc H_{N_c+1}~,\qquad \tilde\pi_{N_c+1}\tmc H_{N_c+1}=\tmc H_{N_c}\otimes\ket{\tilde\omega_{N_c+1}}~,
\eeq
where
\beq
    \ket{\tilde{\omega}_{N_c+1}}\equiv \ket{\omega_{N_c+1}}\otimes\left(\bigotimes_{A=N_c^2}^{(N_c+1)^2-1}\ket{0}^\gh_A\right)\equiv \ket{\omega_{N_c+1}}\otimes\ket{\omega^\gh_{N_c+1}}~.
\eeq
We can write $\tilde\pi_{N_c+1}$ explicitly as
\beq
    \tilde\pi_{N_c+1}=\left(\prod_{i=1}^{N_f}q^\dagger_{N_c+1,i}q_{N_c+1,i}\right)\left(\prod_{j=1}^{N_f}\overline q^{N_c+1\,\dagger}_j\overline q^{N_c+1}_j\right)\left(\prod_{A=N_c^2}^{(N_c+1)^2-1}c^{A\dagger}c^A\right)~.
\eeq
It will be useful to define 
\beq
    \tmc I_{N_c+1}\equiv\left(\tmc H_{N_c}\otimes\ket{\tilde\omega_{N_c+1}}\right)^{\perp}~,
\eeq
as the orthogonal complement of the image of $\tmc H_{N_c}$ under the embedding.
By definition $\tmc I_{N_c+1}$ is the kernel of $\tilde\pi_{N_c+1}$ and it is easy to see that it is spanned by states with at least one of the $q_{N_c+1, i}$, $\overline q^{N_c+1}_i$, or $c^{A>N_c^2}$ occupied.

We now suppose that we have a state $\ket{\tilde\Psi_{N_c}}$ living in the $N_c$-th physical Hilbert space which is identified with a zero ghost cohomology class of $Q_\brst^{(N_c)}$.
We ask when it can be embedded into a physical state of $\tmc H_{N_c+1}$. We first note that the embedding, \eqref{eq:Hkinembed}, induces a pullback on the BRST charge
\beq
    Q^{(N_c+1)}_\brst=Q^{(N_c)}_\brst+\delta_{N_c+1}~,
\eeq
where
\begin{align}
   \delta_{N_c+1}=&\sum_{A=N_c^2}^{(N_c+1)^2-1}c^A\,G_A+\nonumber\\
   &i\left(\sum_{A\geq N_c^2}\sum_{B,C<N_c^2}+2\sum_{B\geq N_c^2}\sum_{A,C<N_c^2}+2\sum_{A,B\geq N_c^2}\sum_{C<N_c^2}+\sum_{A,B,C\geq N_c^2}\right){f^A}_{BC}c^Bc^Cc^{\dagger}_A~. 
\end{align}
Note that because of the ordering prescription of the generators via~\eqref{eq:suNembed1} and~\eqref{eq:suNembed2}, $Q^{(N_c)}_\brst$ does not couple to (anti-)quarks with color index $a=N_c+1$ even when embedded in the covering space.

According to the embedding, a general state in $\tmc H_{N_c+1}$ can be expressed as
\beq\label{eq:PsiNp1decomp}
    \ket{\tilde\Psi_{N_c+1}}=\ket{\tilde\Psi_{N_c}}\otimes\ket{\tilde\omega_{N_c+1}}+\ket{\tilde\alpha_{N_c+1}}~,
\eeq
where $\ket{\tilde\alpha_{N_c+1}}$ lives in $\tmc I_{N_c+1}$, the orthogonal complement to $\tmc H_{N_c}\otimes\ket{\tilde\omega_{N_c+1}}$ in $\tmc H_{N_c+1}$.
We will assume that both $\ket{\tilde\Psi_{N_c}}$ and $\ket{\tilde\Psi_{N_c+1}}$ live in the zero ghost sectors of their respective Hilbert spaces.
Thus if they also correspond to physical states, they are annihilated by $Q^{(N_c)}_\brst$ and $Q^{(N_c+1)}_\brst$, respectively: $Q^{(N_c+1)}\ket{\tilde\Psi_{N_c+1}}=Q^{(N_c)}\ket{\tilde\Psi_{N_c}}=0$.
This implies
\beq\label{eq:embedcriterion}
\delta_{N_c+1}\ket{\tilde\Psi_{N_c}}\otimes\ket{\tilde\omega_{N_c+1}}+Q_\brst^{(N_c+1)}\ket{\tilde\alpha_{N_c+1}}=0~.
\eeq
Consequently, we can rephrase the problem as the following:
a physical state $\ket{\tilde\Psi_{N_c}}\in\tmc H_{N_c}$ (corresponding to a zero ghost cohomology class of $Q_\brst^{(N_c)}$) can be embedded as a physical state $\ket{\tilde\Psi_{N_c+1}}\in\tmc H_{N_c+1}$ if there exists a $\ket{\tilde\alpha_{N_c+1}}\in\tmc I_{N_c+1}$ satisfying~\eqref{eq:embedcriterion}, \textit{i.e.}, $\delta_{N_c+1}\ket{\tilde\Psi_{N_c}}\otimes\ket{\tilde\omega_{N_c+1}}$ is BRST-exact.

We are now able to state concretely our definitions of monotone and fortuitous BRST cohomologies.

\begin{definition}[Monotone and fortuitous states]\label{def:MF} A physical state $\ket{\tilde\Psi_{N_c}}$ is called \textbf{monotone} if it can be embedded as a physical state in $\tmc H_{N_c'}$ for all $N_c'>N_c$ by incrementally implementing~\eqref{eq:PsiNp1decomp} and~\eqref{eq:embedcriterion}.
If not, $\ket{\tilde\Psi_{N_c}}$ is called \textbf{fortuitous}.
\end{definition}

\subsection{Mesons are monotone, baryons are fortuitous}
We will now show that meson states are monotone while baryonic states are fortuitous.
To begin, consider the unnormalized\footnote{This differs from~\eqref{eq:qbmesstate} by the overall coefficient $N_c^{-1/2}$. This obviously doesn't change any conclusions of this section, and simply allows us to define our embedding without the need to carry around and modify factors of $\sqrt{N_c}$.} meson state \beq
\ket{M^{(N_c)}_{ij}}=\sum_{a=1}^{N_c}\bar q_{a,i}q^{a,j}|\tbs{0}\rangle\in\Hph_{N_c}~.
\eeq
We want to find an $\ket{\tilde\alpha_{N_c+1}}$ such that~\eqref{eq:embedcriterion} is satisfied.
This state is easy to find as
\beq
    \ket{\tilde\alpha_{N_c+1}}=\overline q_{N_c+1,i}q^{N_c+1}_{j}|\tbs{0}\rangle~.
\eeq
This is because
\beq
    \ket{M^{(N_c)}_{ij}}\ket{\tilde\omega_{N_c+1}}+\ket{\tilde\alpha_{N_c+1}}=\sum_{a=1}^{N_c+1}\overline q_{a,i}q^a_j|\tbs{0}\rangle~,
\eeq
which is a meson state in $\tmc H_{N_c+1}$ so is annihilated by $Q^{(N_c+1)}$.
This embedding can be done arbitrarily many times (and independent of the number of the flavors) by subsequently adding $\overline q_{N_c+2,i}q^{N_c+2}_j$ and so on.
Thus, by Definition~\ref{def:MF}, the \emph{meson states are monotone} for all $i,j=1,\ldots, N_f$.

Now we turn to baryonic states.
Let $\ket{B^{(N_c)}_{i_1\ldots i_{N_c}}}\in\Hph_{N_c}$; we want to know if we can embed it as a physical state into $\Hph_{N_c+1}$~\eqref{eq:embedcriterion}.
It is clear that the same na\"ive trick for the mesons won't fly here: there is no $\ket{\alpha_{N_c+1}}$ that we can add to $\ket{B_{i_1\ldots i_{N_c}}}\ket{\tilde\omega_{N_c+1}}$ to turn it into a baryon state in $\Hph_{N_c+1}$.
This is because these states have different particle numbers.

While this precludes a baryon state in $\Hph_{N_c+1}$ from being the extension of $\ket{B^{(N_c)}_{i_1\ldots i_{N_c}}}$, we can extend this argument to show that there is no $\ket{\tilde\alpha_{N_c+1}}$ admitting such an extension.
Indeed, since the gauge generators preserve quark particle number,~\eqref{eq:Gkillsn}, then so does both $Q^{(N_c+1)}_\brst$ and $\delta_{N_c+1}$,
\beq\label{eq:Qdeltakillsn}
    [Q^{(N_c+1)}_\brst,\msn]=[\delta_{N_c+1},\msn]=0~.
\eeq
Thus 
\beq
    \msn \delta_{N_c+1}\ket{B_{i_1\ldots i_{N_c}}}\ket{\tilde\omega_{N_c+1}}=N_c\,\delta_{N_c+1}\ket{B_{i_1\ldots i_{N_c}}}\ket{\tilde\omega_{N_c+1}}~.
\eeq
If such an $\ket{\tilde\alpha_{N_c+1}}$ satisfying~\eqref{eq:embedcriterion} existed then it must also have quark number $N_c$.
Thus our putative state
\beq
    \ket{\tilde\Psi_{N_c+1}}=\ket{B_{i_1\ldots i_{N_c}}}\ket{\tilde\omega_{N_c+1}}+\ket{\tilde\alpha_{N_c+1}}~,
\eeq
would be an $N_c$ quark state with zero ghost number and its annihilation by $Q_\brst^{(N_c+1)}$ amounts to finding an $SU(N_c+1)$ invariant combination of $N_c$ quarks.
However, there is no $N_c$ index $SU(N_c+1)$ invariant tensor by which to contract their color indices.
Thus no such state can exist.
We conclude that under Definition~\ref{def:MF}, \emph{baryons are fortuitous}.

Another way to see the obstruction is through the center of $SU(N_c+1)$. The embedded $N_c$-quark baryon transforms with center charge $z^{N_c}$, which is nontrivial for a generator $z\in Z(SU(N_c+1))$. Since $Q_{\brst}^{(N_c+1)}$ preserves quark and anti-quark number, no ghost number-zero correction with the same total quark number can cancel this center charge. Equivalently, the first fundamental theorem for $SL({N_c+1})$ says that an invariant built only from fundamentals requires either pairings with anti-fundamentals or an $(N_c+1)$-index epsilon tensor. Thus the $SU(N_c)$ baryon cannot be in the image of a physical $SU(N_c+1)$ state under the covering projection, $\tilde\pi_{N_c+1}$.

\section{Stabilizer R\'enyi entropy}\label{sec:sre}

We would now like to associate a notion of complexity to states in our qubit quark model and investigate the large $N_c$ scaling of this complexity for the meson and baryon states of the previous section. Let us pause to explain why complexity is of relevance here. In the supersymmetric context, one expects \cite{Chang:2024zqi, Chen:2024oqv} monotone states to be visible via quantization of the gravitational phase space (see \cite{Mikhailov:2000ya, Beasley:2002xv, Lin:2004nb, Berenstein:2004kk, Mandal:2005wv, Grant:2005qc, Maoz:2005nk, Mandal:2006tk, Biswas:2006tj}). This effectively means that the wavefunctions of monotone states are spread over a small $O(1)$ subspace of the full Hilbert space (in some suitable choice of basis) in the $N_c\to\infty$ limit. On the other hand, fortuitous states should constitute the bulk of the exponentially large number of micro-states --- a typical fortuitous state would necessarily be spread out over an exponentially large number of basis states, no matter which basis we pick (unless, of course, we are allowed to pick the basis in a state-dependent manner). In the following, we will try to explore this idea that monotone states should have a small spread in some suitable basis, while fortuitous states should have an exponential spread in our qubit toy model. 

A useful notion of complexity to this end is \emph{stabilizer complexity}, inspired by the resource theory of stabilizer-based quantum computation. In quantum computation, there are a distinguished set of unitary operations (given a choice of computational basis) called the \emph{Clifford} unitaries (to be defined below) and a distinguished set of states, called \emph{stabilizer} states, which can be prepared from the vacuum by Clifford unitaries alone. Clifford unitaries are special because they can be efficiently simulated classically \cite{Veitch:2012ttw,Gottesman:1998hu,Eastin:2009tem} -- they necessarily have polynomial complexity in the number of qubits. And so, the space of stabilizer states can be thought of as a subset of states which is classically simulable. However, Clifford unitaries are not universal, in that one cannot prepare an arbitrary state from these operations alone (or else quantum computation would not be useful) --- one must append additional non-Clifford gates, or equivalently supply non-stabilizer resource states  \cite{Bravyi:2004isx, Chen:2008ixo} to generate all possible unitary operations. Such resource states are referred to as \emph{magic} states, and the resource theory of stabilizer quantum computation \cite{Veitch:2012ttw} tries to quantify the amount of magic, or non-stabilizerness, or non-classicality inherent in a given state. Suitable measures of this non-stabilizerness are known as {\it magic monotones},\footnote{This is not to be equated with the other notion of monotone in this paper: here it is simply refers to a non-increasing property under action of Clifford unitaries.} and we will use a particular one as our proxy for complexity that we will define shortly (see~\cite{White:2020zoz, Cao:2023mzo, Cao:2024nrx, Basu:2024tgg, Basu:2025mmm, Basu:2025uxw, Malvimat:2026oqf, Bettaque:2026vpl} for some recent related work). Intuitively, one can think of a magic monotone as a measure of the complexity of classical simulation with respect to the choice of computational basis. States with $\text{poly}(N_c)$ complexity should be thought of as semiclassical, while those with exponential complexity are inherently quantum. 

\subsection{Review: Stabilizer R\'enyi entropy}

To begin let
\beq
    \mc H=\bigotimes_{\ell=1}^L\,V_\ell
\eeq
be the tensor product of qubit Hilbert spaces (such as those in~\eqref{eq:qbHS}) over some auxiliary label $\ell$. The dimension of $\mc H$ is $2^L$ which we will denote by $\msd$. The collection of Pauli strings 
\beq
    \mathcal P=\Big\{\prod_{\ell=1}^L\sigma^{\mu_\ell}_\ell\Big\}\qquad \text{with}\qquad \sigma^{\mu_\ell}_\ell\in\Big\{\sigma^0\equiv\mathbbm 1,\sigma^x,\sigma^y,\sigma^z\Big\}~,
\eeq
generates all linear operators acting $\mc H$. We define the Clifford group, $\msC$, as the unitary operators acting on $\mc H$ preserving $\mc P$ up to phase valued in ${\pm 1,\pm i}$. That is for 
\beq
    \msC=\Big\{C\in U(\msd)~\Big|~\forall\,P\in\mc P~,\;\;\exists P'\in\mc P~,\;\;\phi\in\mathbb Z_4\;\;\text{s.t.}\;\; C^\dagger P\,C=e^{i\frac{\pi}{2}\phi}P'\Big\}~.
\eeq
The Clifford group generates stabilizer states of qubits:
\beq
    \text{STAB}=\Big\{C\bigotimes_{\ell}\ket{0}_\ell~\Big|~C\in\msC\Big\}~.
\eeq
As stated above, magic are simply states that are not in $\text{STAB}$ and their degree of `non-stabilizerness' is quantified by a magic monotone. In this paper we will focus on a particular monotone\footnote{For pure states and $\alpha\geq 2$, stabilizer entropies can be used as monotones of magic state resource theory. For broader stabilizer protocols including computational-basis measurements, monotonicity and strong monotonicity require care.} called the \emph{stabilizer R\'enyi entropy} (SRE)~\cite{Leone:2021rzd} defined in the following way.

For pure states $\ket{\psi_{1,2}}$ we denote the Pauli average
\beq
    \Xi^{(\alpha)}(\ket{\psi_1},\ket{\psi_2})=\frac{1}{\msd}\sum_{P\in\mc P}\abs{\bra{\psi_1}P\ket{\psi_2}}^{2\alpha}~,
\eeq
and define the SRE as the following
\beq
    \mc M_\alpha\left(\ket{\psi}\right)=\frac{1}{1-\alpha}\log\,\Xi^{(\alpha)}(\ket{\psi},\ket{\psi})~.
\eeq
The SRE satisfies the properties defining a magic monotone:
\begin{itemize}
    \item \textbf{Faithfulness:}
    \beq
        \mc M_{\alpha}\left(\ket{\psi}\right)\geq 0~,\;\;\text{with}\;\; \mc M_\alpha\left(\ket{\psi}\right)=0\;\;\text{iff}\;\;\ket{\psi}\in\text{STAB}~.
    \eeq
    \item \textbf{Clifford invariance:}
    \beq
        \mc M_\alpha\left(C\ket{\psi}\right)=\mc M_\alpha\left(\ket{\psi}\right)\;\;\forall\,C\in \msC~.
    \eeq
    \item \textbf{Additivity:}
    \beq
        \mc M_\alpha\left(\ket{\psi}\otimes\ket{\phi}\right)=\mc M_\alpha\left(\ket{\psi}\right)+\mc M_\alpha\left(\ket{\phi}\right)~.
    \eeq
\end{itemize}

Let us briefly indicate why the SRE, or more specifically, the exponential of the SRE can be treated as a proxy for state complexity. Let $\ket{\phi_\text{res}}$ be a fiducial magic state that we will use as a resource for state preparation. Because the SRE is non-increasing under action of Clifford unitaries, in order to prepare a state $\ket{\psi}$ from $p$ copies of $\ket{\phi_\text{res}}$ (tensored possibly with some number of stabilizer states) using Clifford unitaries alone, it is necessary that
\beq
    \mc M_\alpha\left(\ket{\phi_\text{res}}^{\otimes p}\right)\geq \mc M_{\alpha}(\ket{\psi})~.
\eeq
If the SRE of $\ket{\phi_\text{res}}$ is $\mu^{(\alpha)}$, then we need at least $p\geq \frac{\mc M_\alpha(\ket{\psi})}{\mu^{(\alpha)}}$. In other words the resource needed to prepare $\ket{\psi}$ using Clifford operations alone is at least
\beq
    \ket{\phi_\text{res}}^{\otimes \left(\frac{\mc M_\alpha(\ket{\psi})}{\mu^{(\alpha)}}\right)}~,
\eeq
and thus scales exponentially with $\mc M_\alpha(\ket{\psi})$. To give some intuition in terms of circuit complexity, if the resource $\ket{\phi_\text{res}}$ requires $r$ non-Clifford gates to prepare from $\ket{0}$, then $\ket{\psi}$ requires at least $r^{\mc M_\alpha/\mu^{(\alpha)}}$ such Cliffords (assuming we use $\ket{\phi_\text{res}}$ as the resource). In what follows we will define the stabilizer complexity as
\beq
    \mc C_\alpha=e^{\mc M_\alpha}~.
\eeq

\subsection{SRE in the qubit quark model}

In our qubit quark model we are interested in states in our extended Hilbert space $\tmc H_{N_c,N_f}$ which consists quark, anti-quark, and ghost qubit spaces. However, since all physical states have zero ghost number and since the ghost vacuum has vanishing SRE (this verified explicitly in Appendix~\ref{app:SREdetails}), we only need to consider the contribution of the quarks and anti-quarks. Thus we will have Pauli strings acting each of these spaces, \textit{i.e.}, $\ell$ takes values in $(a,i)$ for the quarks and anti-quarks, and $\msd=\msd_q\,\msd_{\overline q}=2^{2N_fN_c}$. The Pauli operators are given by a Jordan-Wigner transformation of the fermions
\beq\label{eq:JWmap}
    \sigma^x_{(a,i)}=(-1)^{\Gamma_{a,i}}\left(q^a_i+q^\dagger_{a,i}\right)~,\qquad \sigma^y_{(a,i)}=\frac{(-1)^{\Gamma_{a,i}}}{i}\left(q^a_i-q^\dagger_{a,i}\right)~,\qquad \sigma^z_{(a,i)}=q^a_iq_{a,i}^\dagger-q^\dagger_{a,i}q^a_i~.
\eeq
Note that the above indices are \emph{not summed}, \textit{i.e.}, $(a,i)$ are fixed. As standard in the Jordan-Wigner map, the phase, $\Gamma_{a,i}$, is a non-local string 
\beq
    \Gamma_{a,i}:=\sum_{(b,j)<(a,i)}q^b_j\,q^\dagger_{b,j}
\eeq
with respect to some fiducial ordering of the multi-index $(a,i)$ (essentially arranging the quarks along a ``chain'' of flavor and colour indices). While this is necessary to ensure that the Pauli matrices at different ``sites" commute, it will not play an essential role in our computations as will become clear.  Note that these operators are not only gauge-variant, they don't even transform covariantly under $SU(N_c)$. Physical states are sums and products of 
\beq
    \sigma^+_{(a,i)}=\frac{1}{2}(\sigma^x_{(a,i)}+i\sigma^y_{(a,i)})~,
\eeq
acting on the vacuum which will be the ultimate source of the stabilizer complexity. Similar expressions follow for the anti-quarks and we will denote the Pauli operators associated to the anti-quarks with an overbar, \textit{i.e.}, $\bar\sigma^{\bar\mu}$.

\subsubsection*{Meson states}

Using the Jordan-Wigner map, we can write a meson state as
\beq\label{eq:pmesonstate}
    \ket{M_{ij}}=\frac{(-1)^{\gamma_{ij}}}{\sqrt{N_c}}\sum_{a}\sigma^+_{(a,i)}\bar\sigma^+_{(a,j)}\ket{\bs 0,\obs 0}
\eeq
where $(-1)^{\gamma_{ij}}$ is an overall sign coming from pulling the non-local strings to act on the vacuum. The expectation value of a Pauli-string
\beq\label{eq:qbqPauli}
    P=\prod_{(a,i)}\sigma_{(a,i)}^{\mu_{(a,i)}}\bar\sigma^{\bar\mu_{(a,i)}}_{(a,i)}
\eeq
in a meson state can be separated into a diagonal and off-diagonal part
\begin{align}\label{eq:MPM}
    \bra{M_{ij}}P\ket{M_{ij}}=&\frac{1}{N_c}\sum_a\bra{\bs 0,\obs 0}\sigma^-_{(a,i)}\bar\sigma^-_{(a,j)}P\sigma^+_{(a,i)}\bar\sigma^+_{(a,j)}\ket{\bs 0,\obs 0}\nonumber\\
    &\qquad+\frac{2}{N_c}\text{Re}\sum_{a\neq b}\bra{\bs 0,\obs 0}\sigma^-_{(b,i)}\bar\sigma^-_{(b,j)}P\sigma^+_{(a,i)}\bar\sigma^+_{(a,j)}\ket{\bs 0,\obs 0}~.
\end{align}
In Appendix~\ref{app:SREdetails} we evaluate Pauli string expectation values of this type and show that this leads to split of the Pauli average, $\Xi^{(\alpha)}$, into a diagonal and off-diagonal part:
\beq
    \Xi^{(\alpha)}=\Xi^{(\alpha)}_\text{diag.}+\Xi^{(\alpha)}_{\text{off-diag.}}~.
\eeq
Pauli operators contributing to $\Xi^{(\alpha)}_\text{diag.}$ must preserve quark number for each flavor and color index, and so can only consist of $\sigma^{0}$ or $\sigma^z$. On the other hand, the $\Xi^{(\alpha)}_\text{off-diag.}$ necessarily changes quark number between different indices and so receives contributions from Pauli-strings containing $\sigma^x$ and $\sigma^y$. The precise expressions can be found in Appendix~\ref{app:SREdetails}~. In any case it is simple to show that the diagonal part sets an upper bound for the full SRE (see discussion below equation \eqref{eq:Xisplitapp}):
\beq
    \mc M_\alpha^{(\text{diag.})}\equiv\frac{1}{(1-\alpha)}\log \Xi^{(\alpha)}_\text{diag.}\geq \mc M_\alpha~.
\eeq
Let us first focus on the diagonal contribution. Because it only involves Pauli-strings consisting of $\sigma^0$ or $\sigma^z$ (whose vacuum expectation values are 1 or $-1$, respectively) we can express this contribution as the average value of a set of $N_c$ Rademacher random variables\footnote{A Rademacher random variable is a random variable taking values in $\{-1,1\}$ with equal probability.},
\beq\label{eq:expMmeson1}
    \Xi^{(\alpha)}_\text{diag.}=N_c^{-2\alpha}\overline{\abs{\sum_a\chi_{ij}^a}^{2\alpha}}~,\qquad \chi_{ij}^a=\bra{\bs 0,\obs 0}\sigma^{0,z}_{(a,i)}\bar\sigma^{0,z}_{(a,j)}\ket{\bs 0,\obs 0}\in\{-1,1\}~,
\eeq
where the $\overline{\left(\;\ldots\;\right)}$ is the average with $\chi^a_{ij}$ taking values $\pm 1$ with equal probability. The exact expression for this is given by
\begin{align}\label{eq:expMmeson2}
    \Xi^{(\alpha)}_\text{diag.}=&N_c^{-2\alpha}2^{-N_c}\sum_{j=0}^{N_c}\left(\begin{array}{c}N_c\\j\end{array}\right)\abs{N_c-2j}^{2\alpha}\nonumber\\
    =&N_c^{-2\alpha}2^{-N_c}\frac{\dd^{2\alpha}}{\dd t^{2\alpha}}(2\cosh t)^{N_c}\Big|_{t=0}\nonumber\\
    \sim& N_c^{-\alpha}\frac{\Gamma(2\alpha+1)}{2^\alpha \Gamma(\alpha+1)}~.
\end{align}
The exact summation is in the second line performed when $\alpha\in\mathbb Z$ and the expression in the third line is the large $N_c$ behavior which can be analytically continued in $\alpha$. It will be useful to extract this $N_c$ behavior in an alternative manner that will be helpful for the baryon states as well. When $\alpha\in\mathbb Z$ we have
\beq\label{eq:chiavg1}
    \overline{\abs{\sum_a\chi^a_{ij}}^{2\alpha}}=\sum_{a_1,\ldots, a_{2\alpha}=1}^{N_c}\overline{\chi^{a_1}_{ij}\ldots\chi_{ij}^{a_{2\alpha}}}~.
\eeq
Since each $\chi_{ij}^a$ with a fixed color index, $a$, is an independent random variable with vanishing odd moments
\beq
    \overline{\left(\chi^a_{ij}\right)^p}=\frac{1}{2}\left(1+(-1)^p\right)=\delta_{p\text{ even}}~,
\eeq
the only terms in~\eqref{eq:chiavg1} which survive are those when each distinct color index appears an even number of times. Thus~\eqref{eq:expMmeson1} is given by
\beq\label{eq:eMmeson2}
    \Xi^{(\alpha)}_\text{diag.}=N_c^{-2\alpha}\sum_{\{2n_a\}\text{ partition of }2\alpha}\frac{(2\alpha)!}{\prod_{a=1}^{N_c}(2n_a)!}~,
\eeq
\textit{i.e.} $2n_a$ is the number of times the color index $a$ appears in the sum in~\eqref{eq:chiavg1}. $\{2n_a\}$ is an even partition of $2\alpha$ consisting of $N_c$ components. At small values of $\alpha\in\mathbb Z$ we evaluate this expression explicitly as
\begin{align}
    \mc M_2^\text{(diag.)}=&-\log\left(3N_c^{-2}-2N_c^{-3}\right)~,\nonumber\\
    \mc M_3^\text{(diag.)}=&-\frac{1}{2}\log\left(15 N_c^{-3}-30 N_c^{-4}+16 N_c^{-5}\right)~,\nonumber\\
    \mc M_4^\text{(diag.)}=&-\frac{1}{3}\log\left(105 N_c^{-4}-420 N_c^{-5}+588 N_c^{-6}-272 N_c^{-7}\right)~.
\end{align}
The pattern of the large $N_c$ behavior of these expression is evident. At large $N_c$, pair partitions are combinatorially dominant which leads to the scaling
\beq
    \overline{\abs{\sum_a\chi_{ij}^a}^{2\alpha}}\sim (2\alpha-1)!!N_c^{\alpha}~,
\eeq
where $(2\alpha-1)!!=(2\alpha-1)(2\alpha-3)\ldots 1$ is the double factorial. This immediately leads to the second line of~\eqref{eq:expMmeson2}. Since $\mc M_\alpha^\text{(diag.)}$ sets an upper bound on the SRE, this indicates that the large $N_c$ scaling of the SRE for any meson state is at most logarithmic
\beq\label{eq:MmesonlargeN}
    \mc M_\alpha\left(\ket{M_{ij}}\right)\lesssim\frac{\alpha}{\alpha-1}\left(\log N_c+\log c_\alpha\right)
\eeq
where $c_\alpha=2\left(\frac{\Gamma(\alpha+1)}{\Gamma(2\alpha+1)}\right)^{1/\alpha}$ is an $O(1)$ number.

For mesons it is possible for us to exactly evaluate the off-diagonal contribution,~\eqref{eq:mesonODXi} as well,
\beq
    \Xi^{(\alpha)}_\text{off-diag.}=\frac{N_c(N_c-1)}{4}\left(\frac{2}{N_c}\right)^{2\alpha}=2^{2\alpha-2}\left(N_c^{2-2\alpha}-N_c^{1-2\alpha}\right)~,
\eeq
which essentially follows from noting that due to taking $2\text{Re}(\ldots)$ in the off-diagonal part of~\eqref{eq:MPM}, each non-vanishing Pauli expectation value gives a factor of $2^{2\alpha}$ times a multiplicity of keeping an even number of $\sigma^y$'s (we refer the reader to Appendix~\ref{app:SREdetails} for details). Note that for $\alpha=2$ the leading term of $\Xi^{(\alpha)}_\text{off-diag.}$ is the same order as that of $\Xi^{(\alpha)}_\text{diag.}$, \textit{e.g.}, $N_c^{-2}$, however for every $\alpha>2$, $\Xi^{(\alpha)}_\text{off-diag.}$ is subleading to $\Xi^{(\alpha)}_\text{diag.}$ at large $N_c$. Thus the exact $\mc M_\alpha$ retains the logarithmic scaling set by the diagonal upper bound,~\eqref{eq:MmesonlargeN} with only coefficient $c_2$ being modified.

\begin{figure}[h!]
\centering
\includegraphics[width=.8\textwidth]{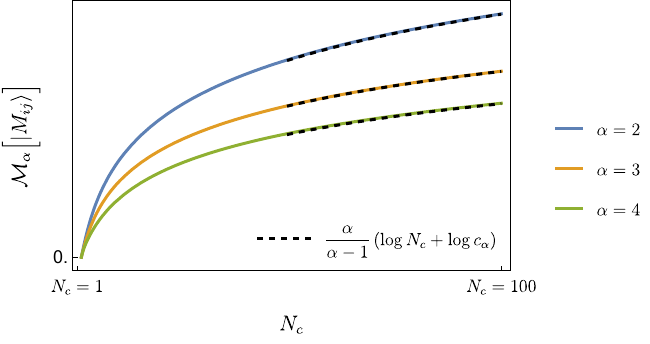}
\caption{The SRE, $\mathcal M_\alpha$, of a meson state, $\ket{M_{ij}}$ for $\alpha=2,3,4$. The solid color lines are the exact evaluation, while the black dashed lines are the large $N_c$ asymptotics (here $c_2=7^{-1/2}$ and $c_{\alpha>2}=2\left(\frac{\Gamma(\alpha+1)}{\Gamma(2\alpha+1)}\right)^{1/\alpha}$).}\label{fig:mesplot}
\end{figure}

In the end we find the exact SRE for the first few values of $\alpha$ is given by 
\begin{align}
    \mc M_2=&-\log\left(7N_c^{-2}-6N_c^{-3}\right)~,\nonumber\\
    \mc M_3=&-\frac{1}{2}\log\left(15 N_c^{-3}-14 N_c^{-4}\right)~,\nonumber\\
    \mc M_4=&-\frac{1}{3}\log\left(105 N_c^{-4}-420 N_c^{-5}+652 N_c^{-6}-336 N_c^{-7}\right)~.
\end{align}
Note that for $N_c=2$, all of the above $
\mc M_\alpha$ vanish, and so the cancellation due to off-diagonal contributions is appreciable. The reason for this cancellation is that the $N_c=2$ state is in fact stabilizer:
\beq
    \ket{M_{ij}}_{N_c=2}=\frac{1}{\sqrt{2}}\left(\ket{1}_{(1,i)}\ket{0}_{(2,i)}\ket{\overline 1}_{(1,j)}\ket{\overline 0}_{(2,j)}+\ket{0}_{(1,i)}\ket{1}_{(2,i)}\ket{\overline 0}_{(1,j)}\ket{\overline 1}_{(2,j)}\right)~.
\eeq
However for large $N_c$ the meson SRE is well approximated by its diagonal contribution. In Figure~\ref{fig:mesplot} we plot the first few $\mc M_\alpha$ both as the exact sum given in~\eqref{eq:expMmeson2} as well as the large $N_c$ asymptotics up to $N_c=100$.

In summary, we have argued that for meson states, the stabilizer complexity $\mathcal{C}_{\alpha}=e^{\mathcal{M}_{\alpha}}$ scales polynomially in $N_c$. This seems to fit well with the expectation that monotone states should be visible in a semiclassical quantization of some large $N_c$, effective phase space. 

\subsubsection*{Baryon states}

Baryon states are comprised entirely of quarks and with the anti-quarks set in the vacuum state, as such their total SRE can be computed from the quark sector alone (the same statement applies {\it mutatis mutandis} to the anti-baryons). A typical baryon state~\eqref{eq:qbbarstate} can be expressed in terms of Pauli operators (up to total sign convention) as
\beq\label{eq:pbarstate}
    \ket{B_{i_1\ldots i_{N_c}}}=\sqrt{\frac{\prod_{i=1}^{N_f}(m_i!)}{N_c!}}\sum_{\varsigma\in\mathfrak S^{\{m_i\}}_{N_c}}\prod_{a=1}^{N_c}\sigma^+_{(a,i_{\varsigma(a)})}\ket{\bs 0}~,
\eeq
where $\mathfrak S^{\{m_i\}}_{N_c}\equiv \mathfrak S_{N_c}/(\mathfrak S_{m_1}\times\ldots\mathfrak S_{m_{N_f}})$ is the permutations of a multiset corresponding to dividing $N_c$ elements into sets of $\{m_i\}$ identical objects. Following similar steps for the meson states (and the computations detailed in Appendix~\ref{app:SREdetails}) the Pauli average splits between diagonal and off-diagonal parts,
\beq
    \Xi^{(\alpha)}[\ket{B_{i_1\ldots i_{N_c}}}]=\Xi^{(\alpha)}_\text{diag.}+\Xi^{(\alpha)}_\text{off-diag.}
\eeq
with the diagonal contribution given by expectation values consisting entirely of $\sigma^0$ and $\sigma^z$. As such, much like in the case of the meson SRE, it can be rewritten as a Rademacher average
\beq\label{eq:eMbaryon2}
    \Xi^{(\alpha)}_\text{diag.}=\left(\frac{\prod_im_i!}{N_c!}\right)^{2\alpha}\overline{\abs{\sum_{\varsigma\in\mathfrak{S}_{N_c}^{\{m_i\}}}\prod_a\Sigma_{a\varsigma(a)}}^{2\alpha}}~.
\eeq
where
\beq
    \Sigma_{ab}:=\bra{\bs 0}_{(a,i_b)}\sigma^{0,z}_{(a,i_b)}\ket{\bs 0}_{(a,i_b)}\in\{\pm 1\}~,
\eeq
is a rectangular $N_c\times \mathsf{M}$ matrix of independent Rademacher variables and $\mathsf M=\sum_{i=1}^{N_f}(1-\delta_{m_i,0})$ is the number of distinct flavor indices in $B_{i_1,\ldots, i_{N_c}}$.
Note that when all of the flavor indices are distinct, $\mathsf{M}=N_c$, then this is the average of the permanent of $\Sigma_{ab}$ raised to the $2\alpha$ power. Due to the various possible index structures both the diagonal and off-diagonal contributions to $\Xi^{(\alpha)}$ are much more intricate for baryons than mesons, and $\Xi^{(\alpha)}_\text{off-diag.}$ in particular is very involved to write down. The exact expression can be found in Appendix~\ref{app:SREdetails}. However, let us first focus on the diagonal contribution, returning to the off-diagonal term afterwards.

For $\alpha\in\mathbb Z$ and a general index set $\{m_i\}$, the diagonal Pauli average,~\eqref{eq:eMbaryon2}, can be expressed as
\beq\label{eq:eMbaryon3}
    \Xi^{(\alpha)}_\text{diag.}=\frac{1}{2^{N_c\mathsf{M}}}\left(\frac{\prod_im_i!}{N_c!}\right)^{2\alpha}\sum_{\{\Sigma_{ab}=\pm1\}}\sum_{\varsigma_1,\ldots\varsigma_{2\alpha}}\prod_{a=1}^{N_c}\prod_{r=1}^{2\alpha}\Sigma_{a,\varsigma_r(a)}~.
\eeq
Because distinct components of $\Sigma_{ab}$ are statistically independent variables with vanishing odd moments, the only surviving terms are when each matrix element $\Sigma_{ab}$ appears an even number of times. Note that this does not necessarily imply that each permutation appears an even number of times, only each index pair $(a,\varsigma_r(a))$ appears an even number of times. Thus the counting problem differs slightly from that of the mesons due to possible combinations of distinct permutations with intersecting images, $\varsigma_r(a)=\varsigma'_r(a)$, that satisfy the parity criterion. 

Instead of evaluating this exactly we can use the properties of Rademacher averages to sandwich $\Xi^{(\alpha)}_\text{diag.}$ between upper and lower bounds. In particular we have
\beq\label{eq:barXidiagbound}
    \left(\frac{N_c!}{\prod_i m_i!}\right)^{-\alpha}\leq \Xi^{(\alpha)}_\text{diag.}\leq  \left(2\alpha-1\right)^{\alpha N_c}\left(\frac{N_c!}{\prod_i m_i!}\right)^{-\alpha}~.
\eeq
The lower bound follows simply from the convexity of $\abs{\ldots}^{2\alpha}$ appearing in the average\footnote{as well as 
\beq
\overline{\left(\sum_{\varsigma}\prod_a\Sigma_{a\varsigma(a)}\right)^2}=\frac{N_c!}{\prod_im_i!}~.
\eeq} of~\eqref{eq:eMbaryon2} and the upper bound follows from a property of Rademacher averages known as hypercontractivity, which we review in Appendix~\ref{app:SREdetails}.

The degree to which~\eqref{eq:barXidiagbound} constrains the diagonal Pauli average in the large $N_c$ limit depends sensitively on the index structure of the baryon in question. For instance, in the extreme case that the baryon is composed completely of identical flavor quarks, $\ket{B_{ii\ldots i}}$, then the only element of $\mathfrak S^{m_i=N_c}_{N_c}$ is the identity with $n_\text{id}=\alpha$. We then find directly from~\eqref{eq:eMbaryon3}
\beq
    \mc M^{(\text{diag.})}_\alpha(\ket{B_{ii\ldots i}})=0\qquad\Longrightarrow \qquad\mc M_\alpha(\ket{B_{ii\ldots i}})=0~.
\eeq
This vanishing of the SRE can be traced to the fact that $\ket{B_{ii\ldots}}=q^1_iq^2_i\ldots q^{N_c}_i\ket{\bs 0}$ is a complete product state in this simple qubit model. As we show in Appendix~\ref{app:SREdetails}, the SRE of any product state of quarks and anti-quarks vanishes. Similarly the lower bound on~\eqref{eq:barXidiagbound} indicates that a baryon with some subset of indices appearing $\sim N_c$ number of times displays at most logarithmic SRE, similar to a meson, \textit{e.g.},
\beq
    \mc M_\alpha\Big(|B_{\,i\underbrace{\scriptstyle j\ldots j}_{N_c-1}}\rangle\Big)\leq\frac{\alpha}{\alpha-1}\log N_c~.
\eeq
Another cautionary example is the baryon with indices appearing a fraction of $N_c$ number of times, \textit{e.g.}, $\ket{B_{i_1\ldots i_1i_2\ldots i_2i_k\ldots i_k}}$ with each index appearing $N_c/k$ has a diagonal contribution bounded by
\beq
    \frac{\alpha}{\alpha-1}\left(N_c\log k-N_c\log(2\alpha-1)\right)\leq\mc M_\alpha^{(\text{diag.})}\leq \frac{\alpha}{\alpha-1}N_c\log k~,
\eeq
and so while could possibly display exponential complexity, the potential corrections are of the same order of magnitude at large $N_c$.

There are a class of baryons however for which~\eqref{eq:barXidiagbound} admits a meaningful large $N_c$ limit: when every flavor index appearing at most a $m_i\sim O(1)$ number of times. An extreme case of this, for instance, is when every flavor index is distinct, $i_1\neq i_2\neq\ldots\neq i_{N_c}$, in which case $\mathfrak{S}^{\{m_i\}}_{N_c}=\mathfrak S_{N_c}$ is just the symmetric group. We will return to this class of baryon states, shortly. For baryons of this type the diagonal contribution of SRE is well approximated as
\beq\label{eq:MbaryonlargeN}
    \mathcal M^\text{(diag.)}_{\alpha}(\ket{B_{i_1,\ldots,i_{N_c}}})\sim\frac{\alpha}{\alpha-1}\left(N_c\log N_c-\sum_{i=1}^{N_f}\log m_i!+O(N_c)\right)~.
\eeq
where, $O(N_c)$ indicates terms at most linear in $N_c$. While we remind the reader that this only serves as an upper bound on the full SRE, $\mc M_\alpha$, it indicates that baryons with flavor indices appearing $m_i\sim O(1)$ number of times may admit super-exponential complexity from the perspective of stabilizer-based computation.\footnote{
The scaling $e^{N_c\log N_c}$ is super-exponential as a function of the rank $N_c$, but in the Veneziano qubit model the number of qubits scales as $n_\text{qubits}\sim 2N_cN_f\sim 2\xi N_c^2$.
Thus the same scaling is $\exp\!\left[O(\sqrt{n_\text{qubits}}\log n_\text{qubits})\right]$, which is sub-exponential in the full qubit system size. 
}

As we have cautioned earlier, $\mc M_\alpha^\text{(diag.)}$ only sets an upper bound on full SRE and so~\eqref{eq:MbaryonlargeN} is not an iron-clad indication of the super-exponential stabilizer complexity of this class of baryon states. It is very possible that much like mesons, the cancellations due to off-diagonal contributions could be appreciable, especially for small $N_c$. This is demonstrably the case for $N_c=2$ for which a baryon with two distinct flavor indices is also stabilizer:
\beq
    \ket{B_{ij}}_{N_c=2}=\frac{1}{\sqrt{2}}\left(\ket{1}_{(1,i)}\ket{0}_{(2,i)}\ket{0}_{(1,j)}\ket{1}_{(2,j)}+\ket{0}_{(1,i)}\ket{1}_{(2,i)}\ket{1}_{(1,j)}\ket{0}_{(2,j)}\right)~.
\eeq
However, one might hope that at large $N_c$, the full SRE is well approximated by~\eqref{eq:MbaryonlargeN} (up to potential additional $O(N_c)$ contributions), at least for baryons with flavor indices appearing $m_i\sim O(1)$ number of times (for which~\eqref{eq:MbaryonlargeN} is well controlled at large $N_c$). At this point we do not have a strong analytic argument for this in full generality. However we can illustrate this is the case in a specific example, namely the baryon composed of entirely distinct flavor indices.

\subsubsection*{Baryons with distinct flavors}

\begin{figure}[h!]
\centering
\includegraphics[width=.8\textwidth]{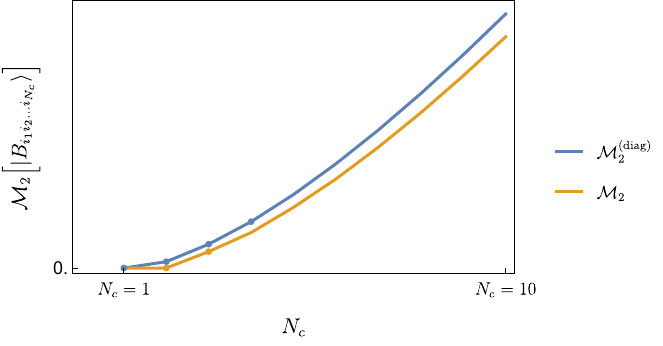}
\includegraphics[width=.8\textwidth]{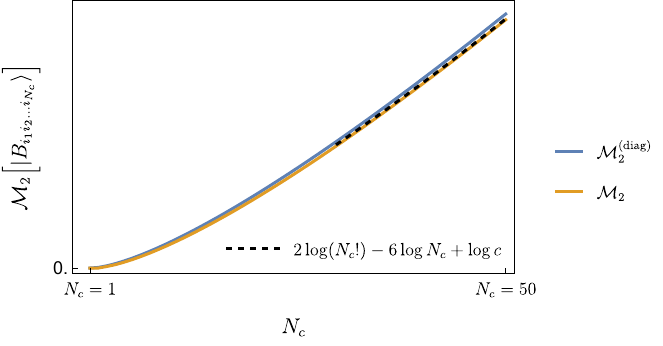}
\caption{The diagonal and full second SRE of the baryon composed of distinct flavor indices. \textbf{(Top)} Comparative plots of based on the formulas~\eqref{eq:diagXi2disbar} and~\eqref{eq:fullXi2disbar} for small $N_c$ ($N_c=1,2,3,4$). The round points are explicit numerical evaluations of the sum over Pauli string expectation values. \textbf{(Bottom)} The large $N_c$ asymptotics; the full second SRE is well approximated by the diagonal contribution. Here $c=e^6720$.}\label{fig:2sre_barplot}
\end{figure}

To illustrate that the diagonal SRE provides a good approximation at large $N_c$ we consider the second SRE, $\mc M_2$, of a baryon state containing entirely distinct flavor indices, $\ket{B_{i_1i_2\ldots i_{N_c}}}$. From~\eqref{eq:eMbaryon3}, the diagonal Pauli average can be written as
\beq
    \Xi^{(2)}_\text{diag.}=\frac{1}{2^{N_c^2}}\frac{1}{(N_c!)^4}\sum_{\{\Sigma_{ab}=\pm1\}}\sum_{\varsigma_1,\ldots,\varsigma_4\in\mathfrak{S}_{N_c}}\prod_{a=1}^{N_c}\Sigma_{a,\varsigma_1(a)}\Sigma_{a,\varsigma_2(a)}\Sigma_{a,\varsigma_3(a)}\Sigma_{a,\varsigma_4(a)}~.
\eeq\
As we explained above, $\Xi^{(2)}_{\text{diag.}}$ simply counts (up to the coefficient $(N_c!)^{-4}$) the number of ways of choosing four permutations, $\{\varsigma_r\}_{r=1,\ldots,4}$, such that each component $\Sigma_{ab}$ appears an even number of times. In Appendix~\ref{app:SREdetails} we show that this counting is reproduced by
\beq\label{eq:diagXi2disbar}
    \Xi^{(2)}_\text{diag.}=\frac{\mathsf{D}_{N_c}}{(N_c!)^2}~,\qquad\qquad \mathsf{D}_{N_c}\equiv\left(\frac{1}{N_c!}\frac{\dd^{N_c}}{\dd t^{N_c}}\frac{e^{-2t}}{(1-t)^3}\right)\Big|_{t=0}\sim \frac{N_c^2}{2e^2}+O(N_c)~.
\eeq
The off-diagonal Pauli average is given by
\beq\label{eq:Xi2barOD1}
    \Xi^{(2)}_\text{off-diag.}=\frac{1}{2^{N_c^2}}\frac{1}{(N_c!)^4}\sum_{\{\mu_{(a,i_b)}\}}\abs{\sum_{\varsigma\neq\tau}\bra{b_\varsigma}\prod_{a,b}\sigma^{\mu_{(a,i_b)}}_{(a,i_b)}\ket{b_\tau}}^4
\eeq
where we introduce the short-hand
\beq
    \ket{b_\sigma}=\prod_{a=1}^{N_c}\sigma^+_{(a,i_{\varsigma(a)})}\ket{\bs 0}~.
\eeq
Sites in which $\varsigma(a)=\tau(a)$ require $\sigma^{\mu_{(a,i_{\varsigma(a)}}}_{(a,i_{\varsigma(a)})}=\sigma^{0,z}$ while sites with $\varsigma(a)\neq\tau(a)$ require either a $\sigma^x$ or $\sigma^y$, acting on both at sites $(a,i_{\varsigma(a)})$ and $(a,i_{\tau(a)})$ in order for a Pauli string to have a non-zero expectation value. Thus~\eqref{eq:Xi2barOD1} breaks up into separate contributions depending on how many colors, $k$, are permuted by the relative permutation, $\vartheta=\varsigma^{-1}\circ\tau$ and the decomposition of $\vartheta$ into disjoint cycles. For each cycle composing $\vartheta$, we must count the allowed assignments of $\sigma^{x,y}$'s in $P$ acting on the sites with $\vartheta(a)\neq a$. The assignment of $\sigma^{0,z}$'s on sites with $a=\vartheta(a)$ mimics the computation of the diagonal contribution with $N_c$ replaced by $N_c-k$. We are then left with summing over $k$ and cycle structures weighted by the number of possible $\vartheta$'s realizing those structures. We perform this counting in detail in Appendix~\ref{app:SREdetails}. The upshot is that full average over Pauli strings (including the diagonal term as the special case, $k=0$) is given by
\beq\label{eq:fullXi2disbar}
    \Xi^{(2)}=\frac{\mathsf{F}_{N_c}}{(N_c!)^2}~,\qquad \mathsf{F}_{N_c}\equiv\left(\frac{1}{N_c!}\frac{\dd^{N_c}}{\dd t^{N_c}}\frac{e^{-6t}}{(1-t)^7}\right)\Big|_{t=0}\sim\frac{N_c^6}{720e^6}+O(N_c^5)~.
\eeq
From~\eqref{eq:diagXi2disbar} and~\eqref{eq:fullXi2disbar} we see that both $\mc M_{2}$ and $\mc M_2^{(\text{diag.})}$ display the same leading behavior at large $N_c$:
\beq
    \mc M_2^{\text{(diag.)}}=2\log(N_c!)-2\log(N_c)+O(1)~,\qquad \mc M_2=2\log(N_c!)-6\log(N_c)+O(1)~,
\eeq
and in particular the super-exponential complexity
\beq
    \mc C_2=e^{2N_c\log N_c+O(\log N_c)}
\eeq
is well captured by the diagonal contribution. In Figure~\ref{fig:2sre_barplot} we check the analytic formulas~\eqref{eq:diagXi2disbar} and~\eqref{eq:fullXi2disbar} against exact numerical evaluation of the SRE for small $N_c$, as well as plot the large $N_c$ behavior.

\subsubsection*{Typical baryons}

One upshot of the previous subsections is that baryons illustrate an interesting subtlety in the matching between fortuity and complexity: while every baryon is fortuitous (as we have shown in Section~\ref{sec:qubit}), not every baryon is super-exponential, or even exponential, in its stabilizer complexity --- it depends on its index structure. However, we can ask what is the stabilizer complexity of a {\it typical} baryon state in the large $N_c$ limit? For instance, when the number of flavors is small, $N_f\sim O(1)$, then typically each flavor index appears an order $N_c$ number of times inside $\ket{B_{i_1\ldots i_{N_c}}}$ resulting in a low diagonal contribution, and thus necessarily a low SRE. On the other hand, in the Veneziano limit in which both $N_f$ and $N_c$ are taken to infinity with $\xi=N_f/N_c$ fixed, then a typical baryon will have an $O(N_c)$ number of distinct indices appearing an $m_i\sim O(1)$ number of times. In this limit a typical baryon will display an upper bound on the SRE that scales super-exponentially as in~\eqref{eq:MbaryonlargeN}.

Let us be a bit more specific. For a given flavor index, $i$, the probability that it appears $k$ times in an flavor symmetric baryon index $(i_1,\ldots,i_{N_c})$, with all other indices unconstrained, is\footnote{More specifically treating each flavor index as a ``bin," then bin $i$ holds $m_i=k$ quanta. Then numerator of~\eqref{eq:pmik} is then how many ways of distributing the remaining $N_c-k$ quanta amongst the other $N_f-1$ bins. The denominator is how many ways to distribute $N_c$ quanta amongst $N_f$ bins.}
\begin{align}\label{eq:pmik}
    p_{m_i=k}&=\frac{\binom{N_c+N_f-k-2}{N_f-2}}{\binom{N_c+N_f-1}{N_f-1}}
    \sim\frac{\xi}{(1+\xi)^{k+1}}~.
\end{align}
We will denote a typical quantity with respect to this probability distribution as $\big<\cdots\big>$. The number of distinct indices appearing $k$ times is
\beq
    \msn_k=N_f\,p_{m_i=k}=\frac{\xi^2N_c}{(1+\xi)^{k+1}}
\eeq
and the total number of distinct non-zero indices appearing in $(i_1,\ldots,i_{N_c})$ is
\beq
    \big<\mathsf M\big>=\sum_{k=1}^{N_c}\msn_k\sim \xi^2 N_c\sum_{k=1}^{\infty}\frac{1}{(1+\xi)^{k+1}}=\frac{\xi}{(1+\xi)}N_c~,
\eeq
each one appearing 
\beq
    \big<m_i\big>=\sum_{k=1}^{N_c}k\,p_{m_i=k}\sim\frac{\xi}{1+\xi}\sum_{k=0}^{\infty}\frac{k}{(1+\xi)^k}=\frac{1}{\xi}~,
\eeq
number of times. As expected, a typical baryon in the Veneziano limit is labelled by $O(N_c)$ distinct indices each appearing $O(1)$ number of times for which~\eqref{eq:MbaryonlargeN} applies. The upper bound for the SRE of a typical baryon in the Veneziano limit is 
\beq
    \big<\mc M^\text{(diag.)}_\alpha\left[\ket{B_{i_1\ldots i_{N_c}}}\right]\big>\sim\frac{\alpha}{\alpha-1}\left(N_c\log N_c-\sum_{k=2}^{N_c}\msn_k\log k!+\log c_\alpha\right)~.
\eeq
Using $\log2\leq\log x!<x^2$ (for $x\geq2$), we sandwich this expression as
\begin{align}
    \frac{\alpha}{\alpha-1}\Big(N_c\log N_c&-\frac{2+\xi}{\xi}N_c+\log c_\alpha\Big)\nonumber\\
    &\lesssim\big<\mc M^\text{(diag.)}_\alpha\left[\ket{B_{i_1\ldots i_{N_c}}}\right]\big>\nonumber\\
    &\qquad\qquad\qquad\lesssim\frac{\alpha}{\alpha-1}\left(N_c\log N_c-\frac{\xi}{(\xi+1)^2}N_c\log 2+\log c_\alpha\right)~.
\end{align}
This gives us strong indication that a typical baryon state has super-exponential stabilizer complexity in Veneziano limit:
\beq
    \mc C_\alpha[\ket{B_{\text{typ}}}]\sim \exp\left(\frac{\alpha}{\alpha-1}N_c\log N_c-\gamma N_c\right)~,\qquad \gamma\sim O(1)~.
\eeq
This resonates with the expectation that typical black hole microstates should have an exponential spread in any fixed basis (such as the Fock basis that we used). 

\section{Discussion and outlook}\label{sec:disc}

In this paper, we have constructed a kinematic generalization of the finite $N$ classification of monotone and fortuitous operators -- originally devised in the context of BPS operators in super Yang-Mills -- in the context of gauge invariant quark operators in a toy model for $SU(N_c)$ QCD with $N_f$ flavors of quarks.
Our construction is based on a covering map (with respect to varying the number of colors) of the BRST cohomology under which mesons appear as monotone operators and baryons appear as fortuitous operators.
We illustrated this distinction in our simplified qubit model of quark operators which captures the basic structure and counting of operators based on $SU(N_c)$ representation theory while allowing an explicit realization of the covering map.
In this qubit quark model we also assigned a notion of complexity to meson and baryon operators associated to the resource theory of stabilizer based computation and with the exponential of the stabilizer R\'enyi entropy as a concrete measure of stabilizer complexity, $\mc C_\alpha =e^{\mc M_\alpha}$.
Under this measure, we showed that meson operators display power law complexity, $\mc C_\alpha\sim N_c^p$, while at large $N_c$ baryons display a myriad of different behaviors of complexity depending on their flavor index structure.
However, in the Veneziano limit when both $N_c$ and $N_f$ are taken to be large, we presented evidence that typical baryons are super-exponentially complex, $\mc C_\alpha\sim e^{N_c\log N_c+O(N_c)}$.

We now elaborate further on these results as well as make some speculative connections between black hole physics and holography.
Before proceeding though, let us be clear that we do not claim that QCD baryons are black hole microstates, nor that BRST fortuity reproduces BPS fortuity.
Rather, baryons provide a simple kinematical model in which rank-dependent gauge invariance, exponential state counting, and large stabilizer complexity appear together.
This is already intriguing.
The analogies between QCD and black hole state counting described in the introduction are also provocative, but these should be regarded only as analogies.

\subsubsection*{On vector vs.\ matrix theories}
The toy model we considered in this paper is vector-like, and it would be interesting to explore generalization to matrix-like theories. (See also~\cite{deMelloKoch:2025ngs,deMelloKoch:2025rkw,deMelloKoch:2026dfo}.) We might expect that the gauge invariant states composed solely of gluon operators might display a sharper connection between fortuity and super-exponential complexity.
This is because gluons are adjoint valued, and so we expect that a fortuitous gluon state utilizing contractions with the $\epsilon^{a_1\ldots a_{N_c}}$ must appear as a determinant operator.
This is morally similar, in terms of counting, to baryon states with all distinct flavor indices.

Instead of having two color indices, the states in our quark model carry a single color index plus a vector index for a large global symmetry group.
We might suspect that our model mimics the physics of other vector models with large global symmetry, such as the $O(N)$ vector models of holography.
Such models display their own interesting finite $N$ effects when restricted to the singlet sector~\cite{Shenker:2011zf,deMelloKoch:2025cec}.
Vector models can also support higher-spin black hole states in three bulk dimensions~\cite{Gutperle:2011kf} and it would be interesting if an appropriate fortuity program can be extended to describe the microstates of such black holes and if they also display a similar relation to complexity as we have described here.

\subsubsection*{On typicality}
Despite the variance in how stabilizer complexity scales in baryon states we have argued that typical baryon states in the Veneziano limit display a super exponential scaling, $\mc C_\alpha\sim e^{N_c\log N_c}$.
This scaling is a bit curious from the usual expectations of typicality: we expect typical states to have support spread over the Hilbert space and thus a complexity scaling like the Hilbert space dimension in this case this is $e^{\log 2\,\xi\,N_c^2}$ (again in the Veneziano limit) \cite{White:2020hgn}.
Of course this discrepancy comes down to an equivocation in the use of ``typical,'' which more standardly refers to a state drawn from the Haar random unitary ensemble.
Here, our notion of typicality is determined from a random draw of a list of baryon flavor indices with equal distribution, \textit{i.e.}, the probability distribution~\eqref{eq:pmik}.
A more appropriate comparison is then to the number of possible baryon states.
However, as we have shown this is only exponential, scaling as $e^{\zeta N_c}$; the stabilizer complexity of a typical baryon scales larger than this.
At present we do not have a strong physical interpretation of the $e^{N_c\log N_c}$ scaling (besides the obvious fact that it arises from a factorial counting).

\subsubsection*{On the differences between QCD, supersymmetric QCD, and our qubit model}
In $\mc N=1$ supersymmetric QCD, there is one nilpotent supercharge, $\overline{\mc Q_{\dot\alpha}}$.
The quarks and anti-quarks are chiral superfields $Q$ and $\widetilde{Q}$, meaning that they are $\overline{\mathcal{Q}}$-closed modulo $\overline{\mathcal{Q}}$-exact.
Chiral gauge invariant operators constructed only from $Q$ and $\widetilde{Q}$ (modulo syzygies and F-term relations from the superpotential~\cite{Gray:2008yu}), form a graded commutative ring --- the chiral ring --- giving holomorphic coordinates on the classical moduli space.
The enumeration of objects in the chiral ring with a specified quark content is read off from coefficients in a Hilbert series which can be calculated via a Molien--Weil integral. In this computation, we do not include anti-chiral superfields $\overline{Q}$ or $\overline{\widetilde{Q}}$, spacetime derivatives (as these produce descendants which are not in the chiral ring), nor the vector multiplet field strength $\mathcal{W}_\alpha$, leading to glueballs.
Because of non-renormalization theorems, the chiral ring is robust (up to known quantum deformations like $\det M - B \widetilde{B} = \Lambda^{2N_c}$ when $N_f = N_c$~\cite{Intriligator:2007cp}).
\comment{
We can think about SQCD from the perspective of geometric invariant theory.
The baryons are Pl\"ucker coordinates, and $q_{i_1} \wedge q_{i_2} \wedge \cdots \wedge q_{i_{N_c}}$ is a section of a determinant line bundle.
The Pl\"ucker embedding is given by the canonical map
\be
\text{Gr}(N_c,N_f) \hookrightarrow \mathbb{P}(\bigwedge{\!}^{N_c} \mathbb{C}^{N_f}) ~.
\ee
That is to say, a quark matrix represents a $N_c$-plane if $\mathbb{C}^{N_f}$, and the wedge of its rows gives a point in $\mathbb{P}(\bigwedge^{N_c} \mathbb{C}^{N_f})$.
The baryons are not independent generators; they satisfy polynomial identities because they come from minors of a single matrix $q$.
The algebra of invariants is therefore generated by mesons and baryons subject to relations.
}

By contrast, in QCD there is no protected cohomological subsector of local gauge invariant operators with analogous properties.
In particular, there is no restriction to holomorphic operators, and operator mixing under renormalization is pervasive: radiative corrections generically mix operators constructed only from quarks and anti-quarks with operators containing derivatives and gluon fields (subject to their quantum numbers).
Counting gauge invariant operators in QCD as well requires tracking the Lorentz structure, $\gamma$-matrix and Fierz identities, equations of motion, integration by parts redundancies, etc.~\cite{Lehman:2015via,Henning:2015daa,Lehman:2015coa,Graf:2020yxt}.
As we stated at the outset, our analysis of states in QCD is restricted to those gauge invariant operators constructed from fermions in the $\bm{N_c}$ and $\overline{\bm{N_c}}$ representations, ignoring the gluons in the adjoint representation and these other intricacies.
The qubit model lies in an intermediate place between SQCD and this simplified model of QCD.

\subsubsection*{On composite gauge invariant operators}
While the details in QCD, SQCD, and our model are different, the color representation theory is the same in all of these theories and is sufficient to recover aspects of the counting of higher-order gauge invariant operators.
The most familiar examples from real world QCD are tetraquarks and pentaquarks, the latter of which have been experimentally observed~\cite{LHCb:2015yax}.
These states also exist in our toy model.
Counting distinct flavor assignments (with no additional Pauli constraints imposed), there are $\sim \frac12N_f^4$ tetraquarks.
These operators can be viewed as a dimeson or diquark--dianti-quark bound states and are monotone.
With gauge group $SU(N_c)$, the analogue of the pentaquarks fall in singlets of $\bm{N_c}^{\otimes (N_c+1)}\otimes \overline{\bm{N_c}}$.
We can think of these as bound states of a baryon and a meson.
A na\"{\i}ve enumeration yields $e^{O(N_c\log N_c)}$ such operators, but imposing a fixed spin / Dirac structure, Pauli statistics correlate the allowed flavor symmetry with the color contraction and reduce the counting.
In the Veneziano limit, we find $e^{\zeta N_c}$ independent operators of this type.
Because the same monomial is not a color singlet of $SU(N_c+1)$ due to the epsilon contraction, these operators are fortuitous.
As we explain in Appendix~\ref{app:GIOs}, any gauge invariant operator constructed from quarks and anti-quarks may be expressed as a linear combination of products of the mesons and baryons.
At large $N_c$ with $\xi := N_f/N_c$ held constant, the number of gauge invariant states associated to fixed meson powers grows polynomially in $N_c$, whereas the number of gauge invariant states in the baryonic sectors grows exponentially.
The former are monotone; the latter are fortuitious.

The consideration of composite operators presents an interesting analogue to the distinction of primary and secondary invariants of~\cite{deMelloKoch:2025ngs,deMelloKoch:2025rkw}.
Namely, while a baryon is an independent gauge invariant operator from a meson, baryons are quadratically reducible, meaning that a product of two baryons can be written in terms of products of mesons (see Appendix~\ref{app:GIOs} for details).
In this language, a baryon is a secondary invariant: it cannot be written in terms of primary invariants (the mesons), but its square can be expressed that way.\footnote{We thank Robert de Mello Koch and Antal Jevicki for discussions on this point.}

\subsubsection*{On baryons and many-body chaos}
Fortuitous operators are intrinsically many-body systems at large $N_c$: a baryon, for instance, contains $N_c$ quarks, so even if each quark--quark interaction is $\mathcal{O}(1/N_c)$ in strength, there are $N_c^2$ pairs, yielding an $\mathcal{O}(N_c)$ net interaction.
Thus, we have a strongly interacting, high density many-body system rather than a simple two-body bound state like a meson.
Moreover, baryons have a rapidly growing number of allowed spin-flavor-orbital configurations; residual interactions mix many nearly degenerate states, so the effective Hamiltonian in a given quantum number sector is a setting where random matrix statistics emerges.
A standard diagnostic is that level spacings of a complicated, non-integrable system show level repulsion and approach Wigner--Dyson statistics (either GOE or GUE depending on symmetries).
The expectation is that sufficiently complicated and highly excited baryon spectra (where many states mix) should fall into this universality class. At large $N_c$, baryons also admit semiclassical descriptions (through either Hartree--Fock or mean field descriptions, or as solitons, \textit{e.g.}, skyrmions).
With enough active degrees of freedom, the corresponding classical collective dynamics can be chaotic, and semiclassical quantization yields Wigner--Dyson-type spectral statistics.
Indeed, random matrices were historically introduced in physics to model spectroscopy and scattering in this setting~\cite{wigner1993characteristic}.

Our qubit model omits much of the above structure.
A more speculative connection to the random statistics inherent to SYK-like systems follows directly from the $SU(N_c)$ invariance itself.
Following Appendix~\ref{app:fabc}, we note that the structure constants of $SU(N_c)$ in~\eref{eq:SUNalg} have familiar statistics in the large $N_c$ limit.
In particular, $\mathbb{E}\!\left[f^{ABC}\right] = 0$ and $\mathbb{E}\!\left[(f^{ABC})^2\right] = N_c^{-3}$.
These are identical scalings to the choice of couplings in a SYK Hamiltonian~\cite{Sachdev_1993}.\footnote{
The structure constants are sparse and basis dependent.
The holographic behavior of the SYK model persists with sparse Hamiltonians~\cite{Xu:2020shn}, and sparseness is not an invariant property under arbitrary Majorana basis rotations.
There are crucial differences, however.
While the quadratic contraction of couplings can reproduce the same large $N$ Schwinger--Dyson structure that gives melonic self-energies, SYK disorder is Gaussian and essentially Wick-factorizing; the structure constants $f^{ABC}$ do not share this property.
The algebra of the $f^{ABC}$s is SYK-like at the level of second moments, but not Gaussian-SYK-like at higher moments.
Indeed, $SU(N)$ $f^{ABC}$ behaves like a sparse deterministic SYK tensor only in observables controlled by $f^{ACD}f^{BCD}\propto\delta^{AB}$.}
This behavior may motivate some previous observations about the role of random matrices in emulating SYK behavior~\cite{Cotler:2016fpe} and explain the similarity to counting of the primary and secondary invariants in~\cite{deMelloKoch:2025ngs,deMelloKoch:2025cec,deMelloKoch:2025rkw,Caputa:2025ikn,deMelloKoch:2026dfo,deMelloKoch:2026utx}.
It also suggests a resemblance to black hole physics.

\subsubsection*{On holographic interpretations}
Witten made the observation that in the $1/N_c$ expansion of QCD, the baryons are solitons in the meson spectrum~\cite{Witten:1979kh}.
This motivates the description of hadrons in the holographic duality between string theory on AdS$_5\times S^5$ and $\mathcal{N}=4$ super-Yang--Mills gauge theory with gauge group $SU(N_c)$~\cite{Maldacena:1997re,Gubser:1998bc,Witten:1998qj}.
A heavy external quark in the fundamental representation is represented by an open fundamental string whose endpoint is localized on the AdS boundary (more precisely, on a probe D$3$/D$7$-brane that reaches the boundary).
A meson is viewed as an open string whose endpoints correspond to a quark and an anti-quark, both at the boundary.
A connecting Wilson line makes the total combination gauge invariant~\cite{Maldacena:1998im,Rey:1998ik}.
That is to say, the physical operator associated to the mesonic string configuration is $q^a(x) W_a^{\,b}(x,y) \overline{q}_b(y)$.
The string can also extend into the AdS bulk and end on a brane.
The worldsheet then describes the quark trajectory, and the brane on which the open string ends is labeled by a Chan--Paton factor, effectively a color index.
On the gravity side, one expects a configuration involving $N_c$ fundamental strings (one per external quark) that together form a color neutral object.

Equivalently, there are $N_c$ units of Ramond--Ramond self-dual five-form flux threading the $S^5$.
A D$5$-brane can wrap this $S^5$ and its worldvolume action contains a Chern--Simons coupling of the form
\be
S_{\text{CS}} \supset \int_{\text{D}5} C_4 \wedge F ~,
\ee
where $C_4$ is the Ramond--Ramond four-form potential with $F_5 = \dd C_4$, and $F$ is the worldvolume field strength on the D$5$-brane.
In the presence of $N_c$ units of $F_5$ flux, this coupling implies that a D$5$-brane wrapping $S^5$ carries $N_c$ units of fundamental string charge.
Gauss's law on the D$5$-brane worldvolume requires that $N_c$ fundamental strings end on the brane in order to neutralize this charge.
The baryon vertex in AdS$_5 \times S^5$ is therefore a D$5$-brane wrapped on $S^5$, which appears as a pointlike object in AdS$_5$, with exactly $N_c$ fundamental strings attached to it~\cite{Witten:1998zw, Witten:1998xy}.
Adding one extra D$3$-brane to the stack whose near-horizon geometry is AdS$_5\times S^5$ introduces an additional unit of flux, so we need to attach one more fundamental string to recover a baryon vertex.
This construction is fortuitous and is a precise structural analogue of the change in trace relations in BPS sectors upon incrementing the rank of the gauge group by one.
 
\comment{
Working in the Poincar\'e patch, the conformal boundary of AdS is at $r\to \infty$.
The baryon vertex (the wrapped D$5$-brane) sits at some finite radial position $r = r_0$ in the bulk.
There are $N_c$ fundamental strings stretched between the baryon vertex and the boundary, with their endpoints at $N_c$ distinct points on the boundary corresponding to the $N_c$ external quarks.
The radial position $r_0$ of the vertex is determined dynamically: the tension of the $N_c$ strings pulling the D$5$-brane toward the boundary is balanced by the D$5$-brane tension and its couplings to the background fields, which tend to keep it deeper in the bulk.
The on-shell configuration of this system computes the expectation value of the baryonic operator in the dual CFT.
}

In the full type IIB supergravity, a D$5$-brane is an extremal black brane solution with a horizon and entropy~\cite{Imamura:1998hf,Brandhuber:1998xy}. 
From the AdS$_5$ point of view, the wrapped D$5$-brane is pointlike, has mass $\sim N_c L/\alpha'$, and carries Ramond--Ramond charge (through $\int_{S^5} F_5$) and fundamental string charge (because $N_c$ strings end on it).
For large $N_c$ and large 't~Hooft coupling $\lambda$, this is a very heavy, compact, charged object.
From sufficiently far away in AdS it resembles a small charged black hole.
This is reminiscent of the Horowitz--Polchinski mechanism~\cite{Horowitz:1996nw}, although here there are some important differences that we point out.
The conformal dimension of the dual operator scales like $\Delta = N_c\sqrt\lambda$.
In supergravity, this black hole has zero horizon area, however, as with the $\frac12$-BPS superstar~\cite{Balasubramanian:2005mg}, we can motivate an entropy in the following way.\footnote{
Operators in the $\frac12$-BPS sector of the $\mathcal{N}=4$ super-Yang--Mills theory in four dimensions are built out of a single chiral superfield $X$ and satisfy the constraint that R-charge equals conformal dimension~\cite{Lin:2004nb}.
Though there are trace relations, the $\frac12$-BPS operators in an $SU(N)$ theory are still $\frac12$-BPS in an $SU(N+1)$ theory.
They are therefore monotone.
The number of such operators with conformal dimension $N^2$ grows like $e^N$, and coarse graining over this ensemble gives a $\frac12$-BPS superstar~\cite{Balasubramanian:2005kk,Balasubramanian:2005mg, Balasubramanian:2018yjq}, which is an incipient black hole~\cite{Myers:2001aq}.
As the entropy is parametrically smaller than $N^2$, the superstar has a horizon with vanishing area in supergravity and looks singular.
This is consistent with the proposal that typical microstates of a black hole with finite horizon in supergravity, such as the $\frac1{16}$-BPS solution, are fortuitous~\cite{Chang:2024zqi,Chen:2024oqv}.}
The baryon vertex can be BPS, but it is only $\frac14$-BPS, and crucially, it is essentially rigid once we fix the external data (how and where the strings end)~\cite{Callan:1998iq,Gomis:1999xs}.
The BPS ``degeneracy'' of a single baryon vertex at fixed charges is expected to be just the finite size of a short $\frac14$-BPS multiplet, \textit{i.e.}, $\mathcal{O}(1)$.
If one adds flavor branes (\textit{e.g.}, probe D$7$-branes) so that the string endpoints carry flavor indices, then baryon operators come in families.
We return to the counting problem we have seen earlier in~\eref{eq:countingbaryons}, namely, choosing $N_c$ fundamentals out of $N_f$ available flavors (with additional structure from spin/R-symmetry and gauge invariant contractions).
If we push this analogy a bit further, one expects a combinatorial factor that yields an entropy
\be
\log d_{\text{flavor}} \sim \zeta N_c
\label{eq:flavor-degeneracy}
\ee
in the Veneziano limit.
In the bulk, the degeneracy is carried by boundary/flavor endpoint degrees of freedom (and D$5$/D$7$ zero modes), not by a macroscopic horizon.
If one relaxes BPS protection and counts generic open string excitations on the F$1$/D$5$ system, the density of states grows exponentially with energy in string units.
Parametrically, for a mass scale $M\sim \Delta/L_\text{AdS}$, one finds
\be
S_\text{string} \sim \frac{M}{T_H} \sim M\ell_s \sim \Delta\,\frac{\ell_s}{L_\text{AdS}}
\sim N_c\sqrt{\lambda}\,\frac{1}{\lambda^{1/4}}
\sim N_c\lambda^{1/4} ~,
\ee
with $T_H$ is the Hagedorn temperature, so that $d\sim \exp(\text{const}\cdot N_c\lambda^{1/4})$.
This reproduces the scaling suggested by a na\"{\i}ve Cardy estimate using a central charge $\sim N_c$ and conformal dimension $\Delta$, even though there is no D$1$/D$5$ system or two-dimensional CFT dual to the baryon vertex.
The physical origin of entropy is stringy in nature.

\section*{Acknowledgments}

We thank Ning Bao, Robert de Mello Koch, Antal Jevicki, and Shiraz Minwalla for comments on a draft of this paper and conversations. 
JRF and VJ both thank TIFR for its hospitality.
JRF additionally acknowledges the hospitality of the Kavli Institute of Physics and Mathematics of the Universe at the University of Tokyo and NYU where part of this work was completed.
JRF is supported by FNRS MISU grant 40024018 ``Pushing Horizons in Black Hole Physics.''
VJ is supported by the South African Research Chairs Initiative of the Department of Science, Technology, and Innovation (DSTI) and the National Research Foundation (NRF), grant 78554.
OP is supported by the Department of Atomic Energy, Government of India, under Project Identification Number RTI-4012 and from the Infosys Endowment for the study of the Quantum Structure of Spacetime.

\appendix
\section{Details on BRST algebra}\label{app:BRST}
The BRST charge is given by~\eqref{eq:BRSTQ}. Using the canonical anti-commutation relations we find the following actions of $Q$ on the quarks and ghosts:
\begin{align}
    \{Q_\brst,q^a_i\}=&c^A{(T_A)^a}_bq^b_i~,\nonumber\\
    \{Q_\brst,q^\dagger_{a,i}\}=&-c_A{(T_A)^b}_aq^\dagger_{b,i}\nonumber\\
    \{Q_\brst,\bar q_{a,i}\}=&-c^A{(T_A)^b}_a\bar q_{b,i}~,\nonumber\\
    \{Q_\brst,\bar q^{\dagger a}_{i}\}=&c_A{(T_A)^a}_b\bar q^{\dagger b}_{i}\nonumber\\
    \{Q_\brst,c^A\}=&\frac{i}{2}{f^A}_{BC}\,c^B\,c^C~,\nonumber\\
    \{Q_\brst,c_A^\dagger\}=&\hat G_A-i{f^B}_{AD}\,c_B^\dagger\,c^D~.
\end{align}
This is easily verified to imply
\beq
    [Q_\brst,\hat G_A]=-i{f^C}_{BA}c^B\,\hat G_C~.
\eeq
It is similarly easily checked that this implies that the BRST action is nilpotent acting on each of these operators:
\beq
    [Q_\brst,\{Q_\brst,q^a_i\}]=[Q_\brst,\{Q_\brst,c^A\}]=[Q_\brst,\{Q_\brst,c_A^\dagger\}]=0~,
\eeq
(nilpotency of the action on $\bar q_{a,i}$ and their conjugates follows similarly) with the final two following from the Jacobi identity of the structure constants:
\beq
    {f^A}_{BC}{f^C}_{DE}+{f^A}_{DC}{f^C}_{EB}+{f^A}_{EC}{f^C}_{BD}=0~.
\eeq
The Hermitian conjugate of the BRST operator is
\beq
    Q^\dagger_\brst=\delta^{AB}c^\dagger_A\hat G_B+\frac{i}{2}\delta_{AD}\delta^{BE}\delta^{CF}{f^A}_{BC}\,c_E^\dagger c_F^\dagger c^D~.
\eeq
The anti-commutator of $Q^\dagger_\brst$ and $Q_\brst$ defines a positive operator akin to a Hamiltonian:
\beq
H_\brst\equiv\{Q^\dagger_\brst,Q_\brst\}
\eeq
This is computed as
\begin{align}
    H_\brst=&\delta^{AB}\hat G_A\hat G_B-i\delta^{AB}{f^C}_{AD}\hat G_B+\frac{1}{4}\delta^{AB}{f^C}_{AD}{f^F}_{BE}c^\dagger_Cc^\dagger_Fc^Dc^E+\frac{N_c}{2}\,c^\dagger_Ac^A\nonumber\\
    =&\delta^{AB}\left(\hat G_A-\frac{i}{2}{f^C}_{AD}c^\dagger_Cc^D\right)\left(\hat G_B-\frac{i}{2}{f^F}_{BE}c_F^\dagger c^E\right) ~,
\end{align}
where the last term the first line comes from the dual Coxeter number of $SU(N_c)$,
\beq
    \delta^{AB}{f^C}_{AD}{f^D}_{BE}=-2N_c\,\delta^C_E ~.
\eeq
The second line makes it clear that $H_\brst$ is the square of a Hermitian operator.

\section{SRE in the qubit quark system}\label{app:SREdetails}

In this appendix we work through the relevant computations of the SRE for the qubit quark model in detail. As a warm-up calculation which will be instructive of the general structure to come, let us first verify that the completely unoccupied state has vanishing SRE. We consider the unoccupied expectation value, $\bra{0}P\ket{0}$, of a general Pauli string
\beq
    P=\prod_\ell\sigma^{\mu_\ell}_\ell ~,
\eeq
for some fixed collection of indices $\{\mu_\ell\}$. Because a $\sigma^x_\ell$ or $\sigma^y_\ell$ acting on $\ket{0}_\ell$ will result in a $\ket{1}_\ell$, this expectation value is non-zero only if every $\sigma_\ell^{\mu_\ell}$ is either $\sigma^0$ or $\sigma^z$, which as 1 or -1 on $\ket{0}_\ell$, respectively. Thus the SRE reduces to a counting (with unit weight) of Pauli-strings consisting only of $\sigma^0$'s and $\sigma^z$'s. Let us denote this subset as $\mc P_{0,z}$:
\beq
    \mc M_\alpha\left(\ket{0}\right)=\frac{1}{1-\alpha}\log\left(\frac{1}{\msd}\sum_{P\in\mc P_{0,z}}\abs{(-1)^{Z_P}}\right)=0~,
\eeq
where $Z_P$ is the number of $\sigma^z$'s contained in $P$ and the last equality following from $\abs{\mc P_{0,z}}=2^L=\msd$.

Now we consider the $n$-quark state formed by the product of $q_\ell$'s 
\beq
    \ket{\psi_A}:=\prod_{\ell\in A}q_\ell\ket{\bs 0}~,\qquad \abs{A}=n~,
\eeq
with respect to some subset of ``sites,'' $A$, and with respect to a fiducial ordering of the $\ell$ indices (in our model $\ell$ would be a multi-index consisting of both the colour and the flavor indices, $\ell=(a,i)$ and this ordering can be taken to be the one defining the phase of the Jordan-Wigner map,~\eqref{eq:JWmap}). Let us compute the overlap of this state with another $n$-quark state associated with a (possible different) set of sites, $B$, with a Pauli-string inserted:
\beq
    \bra{\psi_B}P\ket{\psi_A}=\bra{\bs 0}\left(\prod_{\ell'\in B}^{\leftarrow}q^\dagger_{\ell'}\right)\,P\,\left(\prod_{\ell\in A}q_\ell\right)\ket{\bs 0}~,
\eeq
where the $\leftarrow$ on product over the $q^\dagger$'s indicates that it is taken in the opposite of the fiducial ordering. Under the Jordan-Wigner map,~\eqref{eq:JWmap}, a quark creation operator, $q_\ell$, maps to $\sigma^+_\ell$ while the annihilation operator maps to $\sigma^-_\ell$, where
\beq
    \sigma^\pm=\frac{1}{2}(\sigma^x\pm i\sigma^y)~,
\eeq
up to the non-local string $(-1)^{\Gamma_\ell}$. Pulling this string through to the vacuum results in a pure sign which we will call $(-1)^{\gamma_{A/B}}$. Thus the overlaps is equivalent to 
\begin{align}\label{eq:ABexpvalaspauli}
    \bra{\psi_B}P\ket{\psi_A}=&
    (-1)^{\gamma_A+\gamma_B}\bra{\bs 0}\left(\prod_{\ell'\in B}\sigma^-_{\ell'}\right)P\left(\prod_{\ell\in A}\sigma^+_\ell\right)\ket{\bs 0}~.
\end{align}
We note the commutator of $P$ with a particular $\sigma^+_\ell$ is given by
\beq
    [P,\sigma_\ell^+]=[\sigma^{\mu_\ell}_\ell,\sigma^+_\ell]P_{\setminus\ell}=\left(-\sigma^z_\ell(\delta_{\mu_\ell=x}+i\delta_{\mu_\ell=y})+2\sigma^+_\ell\delta_{\mu_\ell=z}\right)P_{\setminus\ell}~,
\eeq
where $P_{\setminus\ell}$ is the Pauli-string $P$ with $\ell^\text{th}$ site removed and so in total we can write
\begin{align}
    P\sigma^+_\ell=&\left(-\sigma^z_\ell(\delta_{\mu_\ell=x}+i\delta_{\mu_\ell=y})+2\sigma^+_\ell\delta_{\mu_\ell=z}+\sigma^+_\ell\sigma^{\mu_\ell}_\ell\right)P_{\setminus\ell}~.
\end{align}
Note that acting on the vacuum of the $\ell^\text{th}$ side, $\ket{0}_\ell$, the term in parenthesis simplifies to
\beq
    \left(-\sigma^z_\ell(\delta_{\mu_\ell=x}+i\delta_{\mu_\ell=y})+2\sigma^+_\ell\delta_{\mu_\ell=z}+\sigma^+_\ell\sigma^{\mu_\ell}_\ell\right)\ket{0}_\ell=\left(\delta_{\mu_\ell=x}+i\delta_{\mu_\ell=y}+\sigma^+_\ell(\delta_{\mu_\ell=z}+\delta_{\mu_\ell=0})\right)\ket{0}_\ell~.
\eeq
Pulling $P$ through~\eqref{eq:ABexpvalaspauli} results in
\begin{align}
    \bra{\psi_B}&P\ket{\psi_A}\nonumber\\
    &\!\!\!\!\!\!=(-1)^{\gamma_A+\gamma_B}\bra{\bs 0}\left(\prod_{\ell'\in B}\sigma^-_{\ell'}\right)\prod_{\ell\in A}\left(\delta_{\mu_\ell=x}+i\delta_{\mu_\ell=y}+\sigma^+_\ell(\delta_{\mu_\ell=z}+\delta_{\mu_\ell=0})\right)P_{\setminus A}\ket{\bs 0}~,
\end{align}
where $P_{\setminus A}$ is the Pauli-string with all the sites of $A$ removed from $P$. Notice that since both $\ket{\psi_{A/B}}$ are $n$-quark sates, the $\sigma^+$ terms in the $A$ product must pair up with a corresponding $\sigma^-$ term in the $B$ product and so that site must exist in $A\cap B$. For sites in $A\setminus(A\cap B)$ we must keep the identity terms. The remaining terms $\sigma^-$ in $B\setminus(A\cap B)$ must be soaked up by a $\sigma^+$ appearing in the Pauli string, $P\setminus A$. Finally Pauli operators acting on sites not in $A\cup B$ must be either $\sigma^0$ or $\sigma^z$ since they are sandwiched between the vacuum. As a result
\begin{align}
    \bra{\psi_B}P\ket{\psi_A}&=(-1)^{\gamma_A+\gamma_B}(-1)^{Z_{P_{\setminus(A\cup B)}}}\,\delta_{P_{\setminus(A\cup B)}\in\mathcal P_{0,z}}\prod_{\ell\in A\cap B}\left(\delta_{\mu_\ell=z}+\delta_{\mu_\ell=0}\right)\nonumber\\
    &\qquad \qquad \qquad\times\prod_{\ell\in A\setminus(A\cap B)}\left(\delta_{\mu_\ell=x}+i\delta_{\mu_\ell=y}\right)\prod_{\ell\in B\setminus(A\cap B)}\left(\delta_{\mu_\ell=x}-i\delta_{\mu_\ell=y}\right)~,\nonumber\\
    &=(-1)^{\gamma_A+\gamma_B}(-1)^{Z_{P}}\,\delta_{P_{\setminus(A\cup B)}\in\mathcal P_{0,z}}\left[\delta_{P_{A\cap B}\in\mathcal P_{0,z}}\prod_{\ell\in A\cap B}\left(\bra{0}\sigma^{\mu}\ket{0}_\ell\right)\right]\nonumber\\
    &\qquad\qquad\qquad\times\prod_{\ell\in A\setminus(A\cap B)}\left(\delta_{\mu_\ell=x}+i\delta_{\mu_\ell=y}\right)\prod_{\ell\in B\setminus(A\cap B)}\left(\delta_{\mu_\ell=x}-i\delta_{\mu_\ell=y}\right)~.
\end{align}

Note that in the case when $A=B$,
\begin{align}\label{eq:PnquarkEVgen}
    \bra{\psi_A}P\ket{\psi_A}=&(-1)^{Z_P}\delta_{P\in\mathcal P_{0,z}}\bra{\bs 0}\prod_{\ell\in A}\sigma^{\mu_\ell}_{\ell}\ket{\bs 0}~.
\end{align}
As an immediate consequence, we see that the SRE of a quark product state vanishes for similar reasons to that of the vacuum:
\beq
    \mc M_{\alpha}\left(\ket{\psi_A}\right)=0~.
\eeq
It will be relevant for us to consider states that are a superposition over different sets of sites, $\{A_i\}$,
\beq
    \ket{\psi}=\frac{1}{\sqrt{K}}\sum_{i=1}^{K}\ket{\psi_{A_i}}~;
\eeq
for simplicity we will keep all $A_i$ of the same quark number so that the full state is a fixed quark number state. The expectation of a Pauli-string can be written as
\beq
    \bra{\psi}P\ket{\psi}=\frac{1}{K}\left(\sum_i\bra{\psi_{A_i}}P\ket{\psi_{A_i}}+\sum_{i\neq j}\bra{\psi_{A_i}}P\ket{\psi_{A_j}}\right)~.
\eeq
Note that for the terms diagonal sum, $\bra{\psi_{A_i}}P\ket{\psi_{A_i}}$ is necessary that all operators in $P$ are either $\sigma^0$ or $\sigma^z$, while off-diagonal terms must have $\sigma^x$ or $\sigma^y$ operators acting on sites in $A_i\setminus(A_i\cap A_j)$ and $A_j\setminus(A_i\cap A_j)$. Thus the diagonal and off-diagonal terms cannot be simultaneously nonzero. As a result we find the split
\beq\label{eq:Xisplitapp}
    \Xi^{(\alpha)}(\ket{\psi},\ket{\psi})=\Xi^{(\alpha)}_\text{diag}+\Xi^{(\alpha)}_{\text{off-diag}}~,
\eeq
where
\beq
    \Xi^{(\alpha)}_\text{diag}=\frac{1}{\msd}\sum_{P}\abs{\frac{1}{K}\sum_i\bra{\psi_{A_i}}P\ket{\psi_{A_i}}}^{2\alpha}~,\qquad\Xi^{(\alpha)}_\text{off-diag}=\frac{1}{\msd}\sum_{P}\abs{\frac{1}{K}\sum_{i\neq j}\bra{\psi_{A_i}}P\ket{\psi_{A_j}}}^{2\alpha}~.
\eeq
This immediately implies that the SRE is upper-bounded above by the diagonal terms:
\beq
    \mc M_\alpha(\ket{\psi})\leq\mc M_\alpha^{(\text{diag.})}\equiv\frac{1}{1-\alpha}\log\,\Xi^{(\alpha)}_\text{diag}~.
\eeq
It is trivial to extend the above construction to include the anti-quark and the ghost sector. Because the physical states are always have the ghosts in their vacuum and due to the additive property the SRE, the ghost sector never contributes to the SRE of physical states. Thus we will only consider the quarks and anti-quarks. To that end we will write a general Pauli-string as
\beq
    P=\left(\prod_{(a,i)}\sigma^{\mu_{(a,i)}}_{(a,i)}\right)\left(\prod_{(a,i)}\bar\sigma^{\bar\mu_{(a,i)}}_{(a,i)}\right)~.
\eeq
 We then can use~\eqref{eq:PnquarkEVgen}, treating $\ell$ as the multi-index $(a,i)$, to easily evaluate Pauli expectation values in meson and baryon states as in Section~\ref{sec:sre}. 

\subsubsection*{Meson states}

Using the Jordan-Wigner map, we can write a meson state as
\beq\label{eq:pmesonstateapp}
    \ket{M_{ij}}=\frac{(-1)^{\gamma_{ij}}}{\sqrt{N_c}}\sum_{a}\sigma^+_{(a,i)}\bar\sigma^+_{(a,j)}\ket{\bs 0,\obs 0}
\eeq
According to~\eqref{eq:Xisplitapp}, $\Xi^{(\alpha)}$ admits a splitting into a diagonal and off-diagonal part:
\beq
    \Xi^{(\alpha)}=\Xi^{(\alpha)}_\text{diag.}+\Xi^{(\alpha)}_{\text{off-diag.}}
\eeq
where
\beq
    \Xi^{(\alpha)}_\text{diag.}=\frac{2^{2(N_f-1)N_c}}{\msd_q\msd_{\bar q}}\sum_{\{\mu_{(a,i)}=0,z\}}\sum_{\{\bar\mu_{(a,j)}=0,z\}}N_c^{-2\alpha}\abs{\sum_a\bra{\bs 0,\obs 0}\sigma_{(a,i)}^{\mu_{(a,i)}}\bar\sigma^{\bar\mu_{(a,j)}}_{(a,j)}\ket{\bs 0,\obs 0}}^{2\alpha}~,
\eeq
and
\begin{align}\label{eq:mesonODXi}
    \Xi^{(\alpha)}_{\text{off-diag.}}=\frac{2^{2(N_f-1)N_c}}{\msd_q\msd_{\bar q}}\left(\frac{2}{N_c}\right)^{2\alpha}\sum_{\{\mu_{(a,i)}\}}\sum_{\{\bar\mu_{(a,i)}\}}\Big|\sum_{a\neq b}\text{Re}\Big[&\left(\delta_{\mu_{(a,i)}=x}+i\delta_{\mu_{(a,i)}=y}\right)\left(\delta_{\bar\mu_{(a,j)}=x}+i\delta_{\bar\mu_{(a,j)}=y}\right)\nonumber\\
    &\times\left(\delta_{\mu_{(b,i)}=x}-i\delta_{\mu_{(b,i)}=y}\right)\left(\delta_{\bar\mu_{(b,j)}=x}-i\delta_{\bar\mu_{(b,j)}=y}\right)\Big]\nonumber\\
    &\times \delta_{\{\mu\neq\mu_{(a,i)},\mu_{(b,i)}\}={0,z}}\,\delta_{\{\bar\mu\neq\mu_{(a,j)},\bar\mu_{(b,j)}\}={0,z}}\Big|^{2\alpha}
\end{align}

Let us first focus on the contribution from the diagonal part, which can be expressed as the average value of a set of $N_c$ independent random variables,
\beq\label{eq:expMmeson1app}
    \Xi^{(\alpha)}_\text{diag.}=N_c^{-2\alpha}\overline{\abs{\sum_a\chi_{ij}^a}^{2\alpha}}~,\qquad \chi_{ij}^a=\bra{\bs 0,\obs 0}\sigma^{\mu_{(a,i)}}_{(a,i)}\bar\sigma^{\bar\mu_{(a,j)}}_{(a,j)}\ket{\bs 0,\obs 0}\in\{-1,1\}~,
\eeq
where the $\overline{\left(\;\ldots\;\right)}$ is the average with $\chi^a_{ij}$ taking values $\pm 1$ with equal probability. Such random variables are known as Rademacher variables. The exact expression for this is given by
\begin{align}\label{eq:expMmeson2app}
    \Xi^{(\alpha)}_\text{diag.}=&N_c^{-2\alpha}2^{-N_c}\sum_{j=0}^{N_c}\left(\begin{array}{c}N_c\\j\end{array}\right)\abs{N_c-2j}^{2\alpha}=N_c^{-2\alpha}2^{-N_c}\frac{\dd^{2\alpha}}{\dd t^{2\alpha}}(2\cosh t)^{N_c}\Big|_{t=0}~.
\end{align}
The exact summation is in the second line performed when $\alpha\in\mathbb Z$. It will be useful to extract this $N_c$ behavior in an alternative manner applicable to the baryon states. When $\alpha\in\mathbb Z$ we have
\beq\label{eq:chiavg1app}
    \overline{\abs{\sum_a\chi^a_{ij}}^{2\alpha}}=\sum_{a_1,\ldots, a_{2\alpha}=1}^{N_c}\overline{\chi^{a_1}_{ij}\ldots\chi_{ij}^{a_{2\alpha}}}~.
\eeq
Since each $\chi_{ij}^a$ with a fixed color index, $a$, is an independent random variable with vanishing odd moments
\beq
    \overline{\left(\chi^a_{ij}\right)^p}=\frac{1}{2}\left(1+(-1)^p\right)=\delta_{p\text{ even}}~,
\eeq
the only terms in~\eqref{eq:chiavg1app} which survive are those when each distinct color index appears an even number of times. Thus~\eqref{eq:expMmeson1} is given by
\beq\label{eq:eMmeson2app}
    \Xi^{(\alpha)}_\text{diag.}=N_c^{-2\alpha}\sum_{\{2n_a\}\text{ partition of }2\alpha}\frac{(2\alpha)!}{\prod_{a=1}^{N_c}(2n_a)!}~,
\eeq
At large $N_c$, pair partitions are combinatorially dominant which leads to the scaling
\beq
    \Xi^{(\alpha)}_\text{diag}=N_c^{-2\alpha}\overline{\abs{\sum_a\chi_{ij}^a}^{2\alpha}}\sim (2\alpha-1)!!N_c^{-\alpha}~,
\eeq
where $(2\alpha-1)!!=(2\alpha-1)(2\alpha-3)\ldots 1$ is the double factorial. 

We now evaluate the off-diagonal contribution,~\eqref{eq:mesonODXi}.
Separate terms in the sum $\sum_{a\neq b}$ in~\eqref{eq:mesonODXi} have mutually independent delta functions conditions on the Pauli-string. As a result the sum pulls out of the absolute value and simplifies to
\begin{align}
    \Xi^{(\alpha)}_\text{off-diag.}=\frac{1}{16}\left(\frac{2}{N_c}\right)^{2\alpha}\binom{N_c}{2}\sum_{\substack{\mu_{(1,i)} \\ \mu_{(2,i)} }}\sum_{\substack{\bar\mu_{(1,j)} \\ \bar\mu_{(2,j)} }}\Big|\text{Re}\Big[&\left(\delta_{\mu_{(1,i)}=x}+i\delta_{\mu_{(1,i)}=y}\right)\left(\delta_{\bar\mu_{(1,j)}=x}+i\delta_{\bar\mu_{(1,j)}=y}\right)\nonumber\\
    &\times\left(\delta_{\mu_{(2,i)}=x}-i\delta_{\mu_{(2,i)}=y}\right)\left(\delta_{\bar\mu_{(2,j)}=x}-i\delta_{\bar\mu_{(2,j)}=y}\right)\Big]\Big|^{2\alpha}~.
\end{align}
This is only non-zero if an even number of $\sigma^y$'s are chosen, of which there are eight possible choices and so
\beq
    \Xi^{(\alpha)}_\text{off-diag.}=\frac{N_c(N_c-1)}{4}\left(\frac{2}{N_c}\right)^{2\alpha}=2^{2\alpha-2}\left(N_c^{2-2\alpha}-N_c^{1-2\alpha}\right)~.
\eeq
In the end we find the exact SRE for the first few values of $\alpha$ is given by 
\begin{align}
    \mc M_2=&-\log\left(7N_c^{-2}-6N_c^{-3}\right)~,\nonumber\\
    \mc M_3=&-\frac{1}{2}\log\left(15 N_c^{-3}-14 N_c^{-4}\right)~,\nonumber\\
    \mc M_4=&-\frac{1}{3}\log\left(105 N_c^{-4}-420 N_c^{-5}+652 N_c^{-6}-336 N_c^{-7}\right)~.
\end{align}

\subsubsection*{Baryon states}

A generic baryon state~\eqref{eq:qbbarstate} can be expressed in terms of Pauli operators (up to total sign convention) as
\beq\label{eq:pbarstateapp}
    \ket{B_{i_1\ldots i_{N_c}}}=\sqrt{\frac{\prod_{i=1}^{N_f}(m_i!)}{N_c!}}\sum_{\varsigma\in\mathfrak S^{\{m_i\}}_{N_c}}\prod_{a=1}^{N_c}\sigma^+_{(a,i_{\varsigma(a)})}\ket{\bs 0}~,
\eeq
where $\mathfrak S^{\{m_i\}}_{N_c}\equiv \mathfrak S_{N_c}/(\mathfrak S_{m_1}\times\ldots\mathfrak S_{m_{N_f}})$ is the permutations of a multiset corresponding to dividing $N_c$ elements into sets of $\{m_i\}$ identical objects.
Again, $\Xi^{(\alpha)}$ splits between diagonal and off-diagonal parts with
\beq\label{eq:eMbaryon1app}
    \Xi^{(\alpha)}_\text{diag.}=\frac{2^{N_c(N_f-\mathsf{M})}}{\msd_q}\left(\frac{\prod_im_i!}{N_c!}\right)^{2\alpha}\sum_{\{\mu_{(a,i_b)}=0,z\}}\abs{\sum_{\varsigma}\bra{\bs 0}\prod_{a}\sigma^{\mu_{(a,i_{\varsigma(a)})}}_{(a,i_{\varsigma(a)})}\ket{\bs 0}}^{2\alpha}~,
\eeq
where $\mathsf M=\sum_{i=1}^{N_f}(1-\delta_{m_i,0})$ is the number of distinct flavor indices in $B_{i_1,\ldots, i_{N_c}}$ and
\begin{align}\label{eq:eMODbaryonapp}
    \Xi^{(\alpha)}_\text{off-diag.}=\frac{2^{N_c(N_f-\mathsf M)}}{\msd_q}\left(2\frac{\prod_im_i!}{N_c!}\right)^{2\alpha}\sum_{\{\mu_{(a,i_b)}\}}&\left|\sum_{\varsigma}\sum_{\vartheta\neq\text{id}}\Big(\prod_{a|\vartheta(a)=a}\delta_{\mu_{(a,i_{\varsigma(a)})}=0,z}\bra{\bs 0}\sigma^{\mu_{(a,i\varsigma(a))}}_{(a,i_\varsigma(a))}\ket{\bs 0}\Big)\right.\nonumber\\
    &\left.\qquad\qquad\;\;\text{Re}\Big[\prod_{a|\vartheta(a)\neq a}\left(\delta_{\mu_{(a,i_{\varsigma(a)})}=x}+i\delta_{\mu_{(a,i_{\varsigma(a)})}=y}\right)\right.\nonumber\\
    &\qquad\qquad\qquad\qquad \qquad \left.\left(\delta_{\mu_{(a,i_{\varsigma\circ\vartheta(a)})}=x}-i\delta_{\mu_{(a,i_{\varsigma\circ\vartheta(a)})}=y}\right)\Big]\right|^{2\alpha}~.
\end{align}
To evaluate $\Xi^{(\alpha)}_\text{diag.}$ we proceed in a similar way to the mesons. Namely let us denote
\beq
    \Sigma_{ab}:=\bra{\bs 0}_{(a,i_b)}\sigma^{\mu_{(a,i_b)}}_{(a,i_b)}\ket{\bs 0}_{(a,i_b)}\in\{\pm 1\}~,
\eeq
be a rectangular $N_c\times \mathsf{M}$ matrix\footnote{$\Sigma$ depends on the flavor index list, $(i_1,\ldots, i_N)$, since the $i_b$ takes values in the distinct flavor indices appearing in the list. For brevity, we will leave this dependence implicit.} whose components are an independent Rademacher variables. Then the above can be written as
\beq\label{eq:eMbaryon2app}
    \Xi^{(\alpha)}_\text{diag.}=\left(\frac{\prod_im_i!}{N_c!}\right)^{2\alpha}\overline{\abs{\sum_{\varsigma}\prod_a\Sigma_{a\varsigma(a)}}^{2\alpha}}~.
\eeq
Note that when all of the flavor indices are distinct, $\mathsf{M}=N_c$, then this is the average of the permanent of $\Sigma$ raised to the $2\alpha$ power. For $\alpha\in\mathbb Z$,~\eqref{eq:eMbaryon2app} can be expressed as
\beq\label{eq:eMbaryon3app}
    \Xi^{(\alpha)}_\text{diag.}=\frac{1}{2^{N_c\mathsf{M}}}\left(\frac{\prod_im_i!}{N_c!}\right)^{2\alpha}\sum_{\{\Sigma_{ab}=\pm1\}}\sum_{\varsigma_1,\ldots\varsigma_{2\alpha}}\prod_{a=1}^{N_c}\prod_{r=1}^{2\alpha}\Sigma_{a,\varsigma_r(a)}~.
\eeq
Because distinct components of $\Sigma_{ab}$ are statistically independent variables with vanishing odd moments, the only surviving terms are when each index pair $(a,\varsigma_r(a))$ appears an even number of times. This counting is more intricate than the mesons. We can lower and upper bound~\eqref{eq:eMbaryon3app} in the following way. Convexity of $\overline{\abs{\ldots}^{2\alpha}}$ implies that
\beq
    \overline{\abs{\sum_{\varsigma}\prod_a\Sigma_{a\varsigma(a)}}^{2\alpha}}\geq\left(\overline{\abs{\sum_{\varsigma}\prod_a\Sigma_{a\varsigma(a)}}^2}\right)^\alpha=\left(\sum_{\varsigma,\varsigma'}\prod_a\delta_{\varsigma(a),\varsigma'(a)}\right)^\alpha=\left(\frac{N_c!}{\prod_im_i!}\right)^\alpha~.
\eeq
For the upper bound we use the {\it hypercontractivity} property of Rademacher averages which states that for any Rademacher polynomial, $f$, of degree $k$,
\beq
    \rho^k\Big(\overline{f^{\mathsf q}}\Big)^{\frac{1}{\mathsf q}}\leq \Big(\overline{f^{\mathsf p}}\Big)^{\frac{1}{\mathsf p}}~,\qquad 1\leq p\leq q
\eeq
for any $0\leq\rho\leq\sqrt{\frac{\mathsf{p}-1}{\mathsf q-1}}$~\cite{bonami1970etude,beckner1975inequalities}. We note that $\sum_{\varsigma}\prod_a\Sigma_{a\varsigma(a)}$ is Rademacher polynomial of degree $N_c$ and so choosing the $\rho$ for the tightest bound we find
\beq
    \overline{\abs{\sum_{\varsigma}\prod_a\Sigma_{a\varsigma(a)}}^{2\alpha}}\leq (2\alpha-1)^{\alpha N_c}\left(\overline{\abs{\sum_{\varsigma}\prod_a\Sigma_{a\varsigma(a)}}^2}\right)^\alpha=(2\alpha-1)^{\alpha N_c}\left(\frac{N_c!}{\prod_im_i!}\right)^\alpha~.
\eeq
At this point, we do not have an analytic route to estimating the order of magnitude of the off-diagonal contributions in full generality. However we can illustrate the dominance of the diagonal contributions in a specific example.

We consider the second SRE of a baryon state composed of entirely distinct flavor indices $\ket{B_{i_1i_2\ldots i_{N_c}}}$ with $i_1\neq i_2\neq\ldots\neq i_{N_c}$. Let us compute the diagonal contribution first, which from~\eqref{eq:eMbaryon2app} is given by
\beq
    \Xi^{(2)}_\text{diag.}=\frac{1}{(N_c!)^4}\frac{1}{2^{N_c^2}}\sum_{\Sigma_{ab}=\pm1}\sum_{\varsigma_r\in\mathfrak{S}_{N_c}}\prod_a\Sigma_{a,\varsigma_1(a)}\Sigma_{a,\varsigma_2(a)}\Sigma_{a,\varsigma_3(a)}\Sigma_{a,\varsigma_4(a)}~,
\eeq
where $r=1,\ldots,4$. In order for the Rademacher average of $\Sigma_{ab}$ to be non-zero, each index pair $(a,b)$ must appear an even number of times under the images of the permutations $\varsigma_r$. Since the sum over permutation is homogeneous, we can look at the relative permutations with respect to, say $\varsigma_1$, \textit{i.e.}, $\tau_{r>1}:=\varsigma_1^{-1}\circ\varsigma_r$. Let's suppose $\tau_2$ permutes $k$ colors and leaves $(N_c-k)$ colors unpermuted. These $(N_c-k)$ unpermuted index pairs appear twice since $(a,\varsigma_1(a))=(a,\varsigma_2(a))$; the color $a$ may also be permuted by $\tau_3$ and $\tau_4$ but it must also be a shared permutation, $\tau_3(a)=\tau_4(a)$, so that the image appears an even number of times. There are $(N_c-k)!$ such shared permutations. We now consider the colors permuted by $\tau_2$. Decomposing $\tau_2$ into cycles, each cycle must be matched by either $\tau_3$ or $\tau_4$ --- but only one of them --- in order for its image to appear twice. Thus for $c$ cycles, there are $2^c$ choices. Lastly there is an overall factor of $N_c!$ corresponding to the sum over $\varsigma_1$. The Rademacher average reduces to
\beq
    \Xi^{(2)}_\text{diag.}=\frac{1}{(N_c!)^4}\times (N_c!)\sum_{k=0}^{N_c}\sum_{c}\binom{N_c}{k}(N_c-k)!2^cD_{k,c}=\frac{1}{(N_c!)^2}\sum_{k=0}^{N_c}\sum_{c}\frac{D_{k,c}2^c}{k!}~,
\eeq
where $D_{k,c}$ is the number of {\it derangements}\footnote{\textit{i.e.}, a permutation leaving $k$ elements unfixed.} of $k$ objects composed of $c$ cycles.\footnote{For the $k=0$ term of the sum we set by convention $D_{0,0}=1$.} These numbers admit an exponential generating functional~\cite{stanley1986enumerative},
\beq\label{eq:derangEGF}
    \sum_{k=0}^{\infty}\sum_{c}\frac{D_{k,c}x^c\,t^k}{k!}=\frac{e^{-xt}}{(1-t)^x}~,
\eeq
which implies that $\Xi^{(2)}_\text{diag.}$ is given by the coefficient, $\mathsf{D}_{N_c}$, of $t^{N_c}$ in the generating functional\footnote{This follows from~\eqref{eq:derangEGF} by swapping the order of sums:
\beq
    \sum_{N_c=0}^\infty t^{N_c}\sum_{k=0}^{N_c}\sum_{c=1}^k\frac{D_{k,c}x^c}{k!}=\sum_{k=0}^\infty\sum_{c=1}^k\frac{D_{k,c}x^c}{k!}\sum_{N_c\geq k}t^{N_c}=\sum_{k=0}^\infty\sum_{c}\frac{D_{k,c}x^c}{k!}\frac{t^k}{(1-t)}~.
\eeq}
\beq
    \sum_{N_c=0}^{\infty}\mathsf{D}_{N_c}t^{N_c}\equiv\sum_{N_c=0}^{\infty}t^{N_c}\sum_{k=0}^{N_c}\sum_{c}\frac{D_{k,2}2^c}{k!}=\frac{e^{-2t}}{(1-t)^3}~.\label{eq:diagGF}
\eeq
Thus we find
\beq\
    \Xi^{(2)}_\text{diag.}=\frac{1}{(N_c!)^2}\left(\frac{1}{N_c!}\frac{\dd^{N_c}}{\dd t^{N_c}}\frac{e^{-2t}}{(1-t)^3}\right)\Big|_{t=0}~.
\eeq
Now we compute the off-diagonal contribution:
\beq\label{eq:Xi2barOD1app}
    \Xi^{(2)}_\text{off-diag.}=\frac{1}{2^{N_c^2}}\frac{1}{(N_c!)^4}\sum_{P\in\mc P}\sum_{\varsigma_r\neq\tau_r}\Big(\bra{b_{\varsigma_1}}P\ket{b_{\tau_1}}\bra{b_{\varsigma_2}}P\ket{b_{\tau_2}}\bra{b_{\varsigma_3}}P\ket{b_{\tau_3}}\bra{b_{\varsigma_4}}P\ket{b_{\tau_4}}\Big)
\eeq
where
\beq
    \ket{b_\varsigma}=\prod_{a=1}^{N_c}\sigma^+_{(a,\varsigma(a))}\ket{\bs 0}~.
\eeq
Let $\vartheta_r=\varsigma_r^{-1}\circ \tau_r\neq \text{id}$, be the relative permutation of a particular pair appearing in the sums of~\eqref{eq:Xi2barOD1app}. As emphasized by~\eqref{eq:eMODbaryonapp}, for each $\vartheta_r$, the Pauli string, $P$, must have either $\sigma^{0}$ or $\sigma^z$ on sites $(a,i_{\varsigma_r(a)})$ for which $\vartheta_r(a)=a$, and $\sigma^x$ or $\sigma^y$ on the sites $(a,i_{\varsigma_r(a)})$ and $(a,i_{\tau_r(a)})$ when $\vartheta_r(a)\neq a$. Thus terms appearing in the quadruple sum where the $\vartheta_r$'s permute a different number of colors have vanishing expectation value. The non-vanishing Pauli expectation values are further constrained by the cycle decomposition of the relative permutations, $\vartheta_r=\gamma^{(1)}_r\circ\ldots\circ\gamma^{(c_r)}_r$ where $\gamma_r^{(i)}$ are disjoint cycles of length $\ell_r^{(i)}$ (thus $\sum_{i=1}^c\ell_r^{(i)}=k$). We can consider a particular cycle for instance (suppressing unnecessary indices for a moment), $\gamma$ and a color in its domain, $a$. Then $\bra{\varsigma}P\ket{\tau}$ is non-zero only if $P$ flips spins (\textit{i.e.}, has a $\sigma^x$ or $\sigma^y$) on sites $(a,i_{\varsigma(a)})(a,i_{\varsigma\circ\gamma(a)})(\gamma(a),i_{\varsigma\circ\gamma(a)})(\gamma(a),i_{\varsigma\circ\gamma^2(a)})\ldots (\gamma^{\ell-1}(a),i_{\varsigma\circ\gamma^{\ell-1}(a)})(\gamma^{\ell-1}(a),i_{\varsigma(a)})$. That is, the spin-flip structure of contributing Pauli strings breaks up into islands of $2\ell$ sites connected by a cyclic structure and corresponding to a cycle of a relative permutation between a potential pair, $\varsigma$ and $\tau$. If we fix an allowed Pauli string its structure of spin-flip operators will correspond to some cycle structure $\{\gamma^{(i)}_P\}_{i=1,\ldots c_P}$ and $\abs{\bra{B_{i_1\ldots}}P\ket{B_{i_1\ldots}}}^4$ will count how many pairs of permutations $(\varsigma_r,\tau_r)$ have $\vartheta_r=\varsigma_r^{-1}\circ\tau_r$ matching that cycle decomposition, weighted by $(N_c!)^{-4}$ and the phases arising the $\bra{b_{\varsigma_r}}P\ket{b_{\tau_r}}$. For each cycle $\gamma_P^{(i)}$ and a compatible pair of permutations, there are two possible assignments, depending if the cycle structure begins with $(a,i_{\varsigma(a)})$ or $(a,i_{\tau(a)})$. 

Over all cycles, there are then $2^{4c}$ possible assignments of permutation pairs in the quadruple sum with a relative permutation matching its cycle structure. Not all of these assignments survive the Pauli average: if a cycle $\gamma_P^{(i)}$ is assigned a quadruple of pairs $\{(\sigma_r,\tau_r)\}$ with an odd number of $\tau$'s at its starting site\footnote{Because of the alternating structure of the spin-flip sites of $\gamma_P^{(i)}$ this starting site assignment fixes the assignments for the remaining sites of the cycle.} then the average is zero. As an example, if the starting site $(a,\gamma_P^{(i)}(a))$ is assigned three $\tau$'s and one $\sigma$ then the Pauli sum on this site is
\beq
    \left(\langle \sigma^x\sigma^+\rangle_{(a,\gamma_P(a))}\right)^3\langle\sigma^-\sigma^x\rangle_{(a,\gamma_P(a))}+\left(\langle \sigma^y\sigma^+\rangle_{(a,\gamma_P(a))}\right)^3\langle \sigma^-\sigma^y\rangle_{(a,\gamma_P(a))}=1-1=0~.
\eeq
Keeping only even parity assignments, there are $8^c$ possible choices of permutation pairs for a given cycle structure $\{\gamma_P^{(i)}\}_{i=1,\ldots,c}$. After making this assignment, the remaining $(N_c-k)$ colors for which each permutation pair agree, $\varsigma_r(a)=\tau_r(a)$, the problem reduces to the diagonal problem we considered above with $N_c\rightarrow (N_c-k)$.

Now we count how many Pauli strings appear in the sum. As we emphasized above, surviving Pauli strings have spin-flips on sites that can be organized by a structure of cycles acting on $k$ sites. At fixed $k$ and fixed number of cycles, there are
\beq
    \binom{N_c}{k}^2\frac{k!D_{k,c}}{2^c}
\eeq
such configurations giving the same contribution (corresponding to choosing $k$ flavor indices and $k$ color indices involved in the spin-flip sites, a choice of ordered derangement, and a division by the over-counting of cycle orientations).

Noting that the diagonal term can be accounted as a $k=0$ case of the above considerations, the full Pauli average is given by
\begin{align}
    \Xi^{(2)}=&\frac{1}{(N_c!)^4}\sum_{k=0}^{N_c}\sum_{c}\binom{N_c}{k}^2\frac{k!D_{k,c}}{2^c}8^c((N_c-k)!)^2\mathsf{D}_{N_c-k}\nonumber\\
    =&\frac{1}{(N_c!)^2}\sum_{k=0}^{N_c}\sum_c\frac{1}{k!}D_{k,c}\mathsf{D_{N_c-k}}4^c\nonumber\\
    \equiv&\frac{\mathsf{F}_{N_c}}{(N_c!)^2}~.
\end{align}
Lastly given the exponential generating functional of derangements,~\eqref{eq:derangEGF}, and the generating function of the diagonal contribution,~\eqref{eq:diagGF}, we can compute $\mathsf F_{N_c}$ as the $t^{N_c}$ coefficient of the generating function
\begin{align}\label{eq:fullGF}
    \sum_{N_c=0}^{\infty}\mathsf F_{N_c}t^{N_c}=&\sum_{N_c=0}^\infty t^{N_c}\sum_{k=0}^{N_c}\sum_c\frac{4^c}{k!}D_{k,c}\mathsf D_{N_c-k}=\sum_{k=0}^{\infty}\sum_c\frac{4^c}{k!}D_{k,c}t^k\sum_{N_c\geq k}\mathsf D_{N_c-k}t^{N_c-k}\nonumber\\
    =&\frac{e^{-4t}}{(1-t)^4}\frac{e^{-2t}}{(1-t)^3}=\frac{e^{-6t}}{(1-t)^7}~.
\end{align}
The upshot is that the full Pauli average for a baryon state composed of distinct flavor indices is given by
\beq
    \Xi^{(2)}=\frac{1}{(N_c!)^2}\left(\frac{1}{N_c!}\frac{\dd^{N_c}}{\dd t^{N_c}}\frac{e^{-6t}}{(1-t)^7}\right)\Big|_{t=0}~.
\eeq
Comparing to $\Xi^{(2)}_\text{diag}$, while the coefficients $\mathsf D_{N_c}$ and $\mathsf F_{N_c}$ scale as monomials ($N_c^2$ and $N_c^6$, respectively) at large $N_c$, the dominant contribution to the SRE is the inverse square of the factorial $(N_c!)^{-2}$ which is already captured by considering diagonal contribution alone.

\section{Gauge invariant operators}\label{app:GIOs}
The quark sector of QCD allows gauge invariant operators other than single mesons and baryons.
The simplest examples beyond a single meson have quark content $qq\bar q\bar q$.
If one only assigns ordered flavor labels to the two quarks and two anti-quarks, one obtains the na\"{\i}ve count $N_f^4$.
This, however, is not yet the number of independent tetraquark states in the fermionic Fock space.
For $N_c\geq 2$, the color singlet subspace of $(\bm{N_c}\otimes \overline{\bm{N_c}})\otimes(\bm{N_c}\otimes \overline{\bm{N_c}})$
is two-dimensional, corresponding to the two contractions
\begin{equation}
(q_i^a\bar q_{j,a})(q_k^b\bar q_{l,b}) ~,
\qquad
(q_i^a\bar q_{j,b})(q_k^b\bar q_{l,a}) ~. \label{eq:tqcontractions}
\end{equation}
The second contraction is, up to the sign fixed by the chosen fermion ordering convention,
\begin{equation}
(q_i^a\bar q_{j,b})(q_k^b\bar q_{l,a}) = \pm M_{il}M_{kj} ~.
\end{equation}
Thus, the ``crossed'' contraction from~\eref{eq:tqcontractions} is just another degree-two meson product with the anti-flavor labels paired differently.
The independent degree-two mesonic operators are the symmetric products of the $N_f^2$ mesons, and hence are counted by
\begin{equation}
\dim \text{Sym}^2(\bm{N_f}\otimes \overline{\bm{N_f}}) = {N_f^2+1\choose 2} \sim \frac12 N_f^4 ~,
\end{equation}
where $\bm{N_f}$ and $\overline{\bm{N_f}}$ are the fundamental and antifundamental representations of the flavor group. 
This same count is obtained in the diquark--dianti-quark basis.
The tetraquark operators are structurally the same no matter the rank of the gauge group.
Thus, the tetraquark operators are monotone.

The fact that $(\bm{N_c}\otimes\overline{\bm{N_c}})^n$ contains a singlet for every $n$ should not be confused with the statement that higher mesonic powers furnish new primitive generators.
The pure mesonic sector is graded by quark and antiquark number: an $n$-meson product lies in bidegree $(n,n)$, and hence is linearly independent of lower powers by number grading whenever it is nonzero.
However, the color singlets at fixed degree $n$ are spanned
by the permutation contractions
\begin{equation}
I_\sigma=\delta_{a_1}^{b_{\sigma(1)}}\cdots \delta_{a_n}^{b_{\sigma(n)}} ~,
\qquad \sigma\in \mathfrak{S}_n ~.
\end{equation}
These are independent as color tensors in the stable range $n\leq N_c$; for $n>N_c$, finite-rank relations appear, beginning with the antisymmetrizer over $N_c+1$ color indices.
On the other hand, every such $\delta$-contracted invariant is a product of degree one
mesons with permuted flavor labels,
\begin{equation}
q_{i_1}^{a_1}\cdots q_{i_n}^{a_n} \bar q_{j_1,b_1}\cdots \bar q_{j_n,b_n} \delta_{a_1}^{b_{\sigma(1)}}\cdots \delta_{a_n}^{b_{\sigma(n)}} = \pm\prod_{r=1}^n M_{i_r j_{\sigma(r)}} ~.
\end{equation}
Higher powers of mesons give new homogeneous sectors but not new primitive mesonic generators.
In the Veneziano limit with $N_f=\xi N_c$, the fixed-$n$ mesonic sector has dimension
\begin{equation}
\dim \text{Sym}^n(\bm{N_f}\otimes \overline{\bm{N_f}}) = {N_f^2+n-1\choose n}\sim \frac{\xi^{2n}}{n!}N_c^{2n} ~.
\end{equation}
The polynomial behavior applies when $n$ is held fixed.
Once $n$ scales with $N_c$, the enumeration of such states becomes super-polynomial.

As pentaquarks have the same fermionic matter content as a baryon plus a meson in $SU(3)$ QCD, the natural analogue in $SU(N_c)$ QCD is a color singlet operator built from $N_c+1$ quarks and one anti-quark.
Such operators arise from the decomposition of the reducible representation $\bm{N_c}^{\otimes (N_c+1)}\otimes \overline{\bm{N_c}}$.
The multiplicity of the color singlet in this tensor product is $N_c$.
If one performs a na\"{\i}ve count of distinct flavor assignments with $N_f$ flavors (ignoring Pauli constraints from locality and spin), then a na\"{\i}ve enumeration of such gauge invariant operators gives $N_c\,N_f^{N_c+2}$.
In the Veneziano limit, this grows like $e^{N_c\log N_c}$.
When building fully local operators with a fixed spin and Dirac structure, the antisymmetry required by Pauli statistics restricts the allowed flavor symmetries for a given color contraction, which ultimately reduces the overall operator count.
In this case, the number of independent operators grows as $e^{\zeta N_c+\mathcal{O}(\log N_c)}$.
The same monomial $q^{N_c+1}\overline{q}$ is not a color singlet if one changes the gauge group to $SU(N_c+1)$, as the corresponding ``pentaquarks'' would instead involve $N_c+2$ quarks because of the epsilon contraction.
Thus, these operators are fortuitous.

Mesons, baryons, and anti-baryons generate a spanning set, subject to relations, for gauge invariant operators whose constituents are quarks and anti-quarks.
(We ignore gluons in the adjoint representation in this discussion.)
Any color contraction is constructed using the invariant objects $\delta_a^b$, $\epsilon_{a_1,\ldots,a_{N_c}}$, and $\epsilon^{b_1,\ldots,b_{N_c}}$.
If a gauge invariant operator contains $m$ quarks and $n$ anti-quarks, we can organize this as
\be
m = N_c\, r + d ~, \qquad n = N_c\, s + d ~,
\ee
because we must use $d$ deltas, $r$ epsilons, and $s$ dual epsilons to contract the color indices.
A gauge invariant operator is chargeless, so
\be
m - n = N_c\, (r - s) \equiv 0\!\!\! \mod N_c ~.
\ee
We have the identity
\be
\epsilon_{a_1,\ldots,a_{N_c}} \epsilon^{b_1,\ldots,b_{N_c}} = \sum_{\sigma\in \mathfrak{S}_{N_c}} \text{sgn}(\sigma) \delta_{a_1}^{b_{\sigma(1)}} \cdots \delta_{a_{N_c}}^{b_{\sigma({N_c})}} ~, \label{eq:epdel}
\ee
where $\sigma$ is a permutation of the indices.
Thus, we can marry $\epsilon_{a_1\ldots a_{N_c}}$ with a dual $\epsilon^{b_1\ldots b_{N_c}}$ and replace the product with (sums of) $N_c$ Kronecker deltas.
So $m = n + k\, N_c$, where $k$ counts the number of unpaired epsilon symbols, meaning that everything is a linear combination of $k$ baryons and $n$ mesons or $k$ anti-baryons and $m$ mesons.
This is the first fundamental theorem for $SL(N)$ stated in QCD language.

Suppose, using~\eref{eq:baryon} and~\eref{eq:anti-baryon}, we apply~\eref{eq:epdel} to express a baryon and an anti-baryon pair in terms of mesons.
The left hand side is fortuitous, and from~\eref{eq:countingbaryons}, we see that
\be
\text{Number of baryon--anti-baryon pairs} = {N_f + N_c - 1 \choose N_c}^2 ~. \label{eq:countingpairs}
\ee
Notice that the number of indices in~\eref{eq:epdel} changes as we increment $N_c$, so the combination on the right hand side will be fortuitous even though the mesons are monotone.
To verify the counting in terms of the mesons, we observe that~\eref{eq:epdel} when contracted with quarks $q^{a}_i$ and anti-quarks $\overline{q}_{j,b}$ is symmetric in the flavor labels $(i_1,\ldots,i_{N_c})$ and $(j_1,\ldots,j_{N_c})$, so the ordering does not matter.
We only care about the occupation numbers
\be
\ell_i = \Big|\{k ~:~ i_k = i\}\Big| ~, \quad \overline{\ell}_j = \Big|\{k ~:~ j_k = j\}\Big| ~, \qquad \sum_i \ell_i = \sum_j \overline{\ell}_j = N_c ~.
\ee
This is a weak composition of $N_c$ into $N_f$ parts, and enumerating the number of choices for $\ell$ and $\overline{\ell}$ precisely recovers the expression in~\eref{eq:countingpairs}.
The scaling of the right hand side is not $N_c! (N_f)^{2N_c}$ as one might na\"{\i}vely expect from the sum over the symmetric group: the $N_c!$ terms in the sum are not distinct, independent states; rather, they are the expansion of a symmetrized operator.
We could equivalently count the $\mathfrak{S}_{N_c}$ orbits of flavor assignments that the sum projects onto in order to obtain~\eref{eq:countingpairs}.
In the Veneziano limit, the scaling of baryon--anti-baryon pairs is $e^{2\zeta N_c}$.
Because of~\eref{eq:epdel} and Grassmann--Pl\"ucker relations, the mesons and baryons constitute an overcomplete basis for the gauge invariant operators.

\section{Structure constants of $SU(N)$}\label{app:fabc}
Let $\{T_A\}_{A=1}^{N^2-1}$ be a basis of the $\mathfrak{su}(N)$ Lie algebra in the fundamental representation, normalized as
\be
\mathrm{tr}(T^A T^B)=\frac12\,\delta^{AB} ~. \label{eq:tt}
\ee
Define the antisymmetric and symmetric invariant tensors $f^{ABC}$ and $d^{ABC}$ by
\be
[T^A,T^B]= i f^{ABC}T^C ~,
\qquad
\{T^A,T^B\}=\frac1N\delta^{AB}\mathbb{I}+ d^{ABC}T^C ~.
\ee
Since $f^{ABC}$ is totally antisymmetric, $\mathbb{E}\!\left[f^{ABC}\right] = 0$.
Though $d^{ABC}$ is totally symmetric, $\mathbb{E}\!\left[d^{ABC}\right]$ also vanishes because there is a sign flip symmetry that preserves~\eref{eq:tt} while sending $T^A\to -T^A$.
Under this flip, $d^{ABC}\to -d^{ABC}$, so averaging over a symmetric ensemble (random indices or random basis) kills this mean.

Two standard identities (valid for the normalization above) are
\be
f^{ACD}f^{BCD}=N\,\delta^{AB} ~,
\qquad
d^{ACD}d^{BCD}=\frac{N^2-4}{N}\,\delta^{AB} ~.
\label{eq:ff-dd}
\ee
Summing~\eref{eq:ff-dd} over $A=B$ yields the total squared norms
\be
\sum_{A,B,C}(f^{ABC})^2 = N(N^2-1) ~,
\qquad
\sum_{A,B,C}(d^{ABC})^2 = \frac{(N^2-4)(N^2-1)}{N} ~.
\label{eq:total-norms}
\ee
These identities already determine the variance scaling of typical components at large $N$.
If we choose a triple $(A,B,C)$ uniformly from $\{1,\dots,N^2-1\}^3$, using~\eref{eq:total-norms},
\be
\mathbb{E}\!\left[(f^{ABC})^2\right] \approx \frac{\sum_{A,B,C}(f^{ABC})^2}{(N^2-1)^3} = \frac{N(N^2-1)}{(N^2-1)^3} \sim \frac{1}{N^3} ~,
\ee
and similarly
\be
\mathbb{E}\!\left[(d^{ABC})^2\right] \approx \frac{\sum_{A,B,C}(d^{ABC})^2}{(N^2-1)^3} \sim \frac{1}{N^3} ~.
\ee
A typical component has magnitude
\be
f^{ABC}\sim N^{-3/2} ~,
\qquad
d^{ABC}\sim N^{-3/2} ~,
\label{eq:typical-scale}
\ee
where typicality is defined by uniformly random adjoint indices.
The structure constants are not independent random variables, however.
They satisfy exact algebraic constraints, most notably the Jacobi identity:
\be
f^{ABE}f^{CDE}+f^{BCE}f^{ADE}+f^{CAE}f^{BDE}=0 ~,
\label{eq:jacobi}
\ee
as well as additional relations involving $d^{ABC}$.

\bibliographystyle{JHEP}
\bibliography{fortSYK.bib}

\end{document}